\documentclass[twocolumn]{aastex62}

\usepackage{amsmath}

\graphicspath{{./}{figures/}}

\shorttitle{PHIBSS2: Field CO Luminosity Functions}
\shortauthors{Lenki\'{c}, Bolatto, F\"orster Schreiber, et al.}

\begin{document}

\title{Plateau de Bure High-z Blue-Sequence Survey 2 (PHIBSS2): Search for Secondary Sources, CO Luminosity Functions in the Field, and the Evolution of Molecular Gas Density through Cosmic Time\footnote{Based on observations of an IRAM Legacy Program carried out with the
NOEMA, operated by the Institute for Radio Astronomy in the Millimetre
Range (IRAM), which is funded by a partnership of INSU/CNRS (France),
MPG (Germany), and IGN (Spain).}}

\correspondingauthor{Laura Lenki\'{c}}
\email{llenkic@astro.umd.edu}

\author[0000-0003-4023-8657]{Laura Lenki\'{c}}
\affil{University of Maryland 1113 PSC Bldg., 415 College Park, MD  20742-0001 }

\author[0000-0002-5480-5686]{Alberto D. Bolatto}
\affiliation{University of Maryland 1113 PSC Bldg., 415 College Park, MD  20742-0001 }

\author[0000-0003-4264-3381]{Natascha M. F\"{o}rster Schreiber}
\affiliation{Max-Planck-Institut f\"{u}r extraterrestrische Physik (MPE), Giessenbachstr., D-85748 Garching, FRG, Germany}

\author[0000-0002-1485-9401]{Linda J. Tacconi}
\affiliation{Max-Planck-Institut f\"{u}r extraterrestrische Physik (MPE), Giessenbachstr., D-85748 Garching, FRG, Germany}

\author[0000-0002-7176-4046]{Roberto Neri}
\affiliation{IRAM, 300 Rue de la Piscine, F-38406 St. Martin d’Heres, Grenoble, France}

\author[0000-0003-2658-7893]{Francoise Combes}
\affiliation{Observatoire de Paris, LERMA, College de France, CNRS, PSL Univ., Sorbonne Univ. UPMC, F-75014, Paris, France}

\author[0000-0003-4793-7880]{Fabian Walter}
\affiliation{Max Planck Institut f\"{u}r Astronomie, K\"{o}nigstuhl 17, 69117 Heidelberg, Germany}
\affiliation{National Radio Astronomy Observatory, Pete V. Domenici Array Science Center, P.O. Box O, Socorro, NM 87801, USA}

\author[0000-0003-0444-6897]{Santiago Garc\'{i}a-Burillo}
\affiliation{Observatorio Astronómico Nacional-OAN, Observatorio de Madrid, Alfonso XII, 3, E-28014—Madrid, Spain}

\author[0000-0002-2767-9653]{Reinhard Genzel}
\affiliation{Max-Planck-Institut f\"{u}r extraterrestrische Physik (MPE), Giessenbachstr., D-85748 Garching, FRG, Germany}
\affiliation{Department of Physics, Le Conte Hall, University of California, Berkeley, CA 94720, USA}
\affiliation{Department of Astronomy, Campbell Hall, University of California, Berkeley, CA 94720, USA}

\author[0000-0003-0291-9582]{Dieter Lutz}
\affiliation{Max-Planck-Institut f\"{u}r extraterrestrische Physik (MPE), Giessenbachstr., D-85748 Garching, FRG, Germany} 

\author[0000-0003-1371-6019]{Michael C. Cooper}
\affiliation{Deptartment of Physics \& Astronomy, Frederick Reines Hall, University of California, Irvine, CA 92697, USA}


\begin{abstract}
We report on the results of a search for serendipitous sources in CO emission in 110 cubes targeting CO\,($2-1$), CO\,($3-2$), and CO\,($6-5$) at $z\sim1-2$ from the second Plateau de Bure High-z Blue-Sequence Survey (PHIBSS2). The PHIBSS2 observations were part of a 4-year legacy program at the IRAM Plateau de Bure Interferometer aimed at studying early galaxy evolution from the perspective of molecular gas reservoirs. We present a catalog of 67 candidate secondary sources from this search, with 45 out of the 110 data cubes showing sources in addition to the primary target that appear to be field detections, unrelated to the central sources. This catalog includes the redshifts, line widths, fluxes, as well as an estimation of their reliability based on their false positive probability. We perform a search in the 3D-HST/CANDELS catalogs for the secondary CO detections and tentatively find that $\sim$ 64\% of these have optical counterparts, which we use to constrain their redshifts. Finally, we use our catalog of candidate CO detections to derive the CO\,($2-1$), CO\,($3-2$), CO\,($4-3$), CO\,($5-4$), and CO\,($6-5$) luminosity functions over a range of redshifts, as well as the molecular gas mass density evolution. Despite the different methodology, these results are in very good agreement with previous observational constraints derived from blind searches in deep fields. They provide an example of the type of ``deep field'' science that can be carried out with targeted observations.
\end{abstract}


\section{Introduction} \label{sec:intro}
Detailed measurements of the star formation history of the universe reveal that the process of galaxy assembly peaked about 10 billion years ago. The star formation rate (SFR) density in galaxies (i.e., total SFR in galaxies in a comoving volume of the universe) across cosmic time is observed to gradually increase to redshifts of $z \gtrsim 2$, peak at $z \sim 1-2$, and then decrease from redshifts of $z \sim 1$ to the present day by almost an order of magnitude \citep[see e.g.,][]{madau14}. The fundamental physical processes that shape this evolution, however, are still uncertain. This evolution may be driven by the availability of larger reservoirs of cold dense molecular gas (the immediate fuel for star formation) in high-$z$ galaxies, by higher efficiencies for converting molecular gas into stars, or by a combination of both. Therefore, it is  interesting to constrain the molecular gas content of galaxies over cosmic time (measured as total gas mass per co-moving volume) in order to understand the evolution of the cosmic star formation history. 

Most studies of the cold molecular gas in galaxies have used CO observations, the most common molecular gas mass tracer \citep{bolatto13}, of galaxies that were pre-selected based on optical or near-infrared surveys. Other galaxies detected in CO at higher redshifts were initially selected from rest-frame far-infrared continuum surveys as sub-mm galaxies \citep{blain02,casey14}. These studies have shaped our understanding of the relation between molecular gas content and star formation in known populations of galaxies. Targeted CO studies find that $z \sim 2$ galaxies have much larger molecular gas reservoirs than local galaxies \citep{greve05,daddi10,genzel10,genzel15,tacconi10,tacconi13,tacconi18,freundlich19} and that the changes in growth history are largely driven by the cold molecular gas mass properties of galaxies. While these types of studies allow us to understand the properties of galaxy samples, they can potentially introduce unknown systematic biases through selection effects. It is therefore beneficial to complement them with blind searches for CO-emitting galaxies.

Spectral scans on specific deep fields have been used to carry out blind searches targeting rotational transitions of CO over wide frequency and redshift ranges, measuring the CO luminosity function at different epochs in the history of the universe. The CO luminosity function so obtained gives a measurement of the molecular gas mass density over the range of redshifts sampled by the observations \citep{carilli13}. Initial efforts that have followed this strategy are the IRAM Plateau de Bure Interferometer (PdBI) observations in the \textit{Hubble} Deep Field North, and the Atacama Large Millimeter Array (ALMA) observations in the \textit{Hubble} Ultra Deep Field (the ASPECS-Pilot program). These spectral scans were conducted at 3 mm and 1 mm wavelengths, covering areas of $\sim 0.5$ and $\sim 1$ arcmin$^{2}$ in size, respectively \citep[see][for survey descriptions]{decarli14,walter16}. \citet{walter14} and \citet{decarli16} present luminosity function measurements for CO\,($3-2$) and higher-J transitions at $z \sim 2-3$, and CO\,($5-4$) and higher-J transitions at $z \sim 5-7$. These studies provided some of the first constraints on the cosmological CO luminosity function, and the cosmic molecular gas mass density evolution, but they are limited by the small areas covered.

More recently, the COLDz project ($>300$ hours of observations on the JVLA) covered a $\sim 48$ armin$^{2}$ area in GOODS-N and a $\sim 8$ arcmin$^{2}$ area in COSMOS in the $30-38$ GHz frequency range, targeting CO\,($1-0$) at $z \sim 2-2.8$ and CO\,($2-1$) at $z \sim 4.9-6.7$ \citep{pavesi18}. This survey provides constraints for the CO luminosity function at $z > 2$ \citep{riechers19}. The ASPECS Large Program (LP; 150 hours of observations on ALMA) covers most of the \textit{Hubble} eXtremely Deep Field ($\sim 4.6$ arcmin$^{2}$) at 3 mm and 1.2 mm wavelengths \citep{gonzalez-lopez19}. \citet{decarli19} use it to measure the CO luminosity function and find that the cosmic molecular gas mass density peaks at $z \sim 1.5$ and decreases by a factor of $\sim 6.5^{+1.8}_{-1.4}$ to the present day.

In this paper, we present CO luminosity function and cosmic molecular gas mass density evolution measurements we make from repurposed data which takes advantage of independent deep observations of targeted galaxies. Specifically, we present the results from a ``blind" CO search in the second Plateau de Bure High-z Blue-Sequence Survey (PHIBSS2) observations, the follow up to PHIBSS. PHIBSS and PHIBSS2 have been productive surveys with key results on their main objective, characterizing normal $z\sim1-2$ galaxies. Among other results, PHIBSS and PHIBSS2 have yielded scaling relations for main sequence galaxies at those redshifts, depletion times and molecular fractions \citep{genzel15,tacconi18}, and characterized molecular reservoirs for $z<1$ galaxies \citep{freundlich19}. However, these observations also have the potential to yield impactful ``deep field'' science. Since each PHIBSS2 observation targeted a galaxy selected from the 3D-HST CANDELS fields, characterization of serendipitous detections benefits from the extensive multi-wavelength coverage available in these legacy fields.

This paper is structured as follows: Section \ref{sec:obs} summarizes the observations used, Section \ref{sec:method} describes the blind search algorithm, the optical counterpart search, and our statistical methods for assessing the likelihood that each candidate is real as well as the completeness of the search algorithm. Section \ref{sec:results} presents the results of the line search, the CO luminosity functions we derive, and the molecular gas mass density evolution constraints. Sections \ref{sec:discussion} compares to previous works and Section \ref{sec:conclusions} summarizes the work done. The properties of the candidate sources, their spectra, and optical counterparts are presented in the Appendices.

Throughout the paper, we assume $\Lambda$CDM cosmology with $H_{0} = 70$ km\,s$^{-1}$\,Mpc$^{-1}$, $\Omega_{m} = 0.3$, and $\Omega_{\Lambda} = 0.7$, consistent with the \textit{Wilkinson Microwave Anisotropy Probe} measurements \citep{komatsu11}.

\section{Observations} \label{sec:obs}
\subsection{The ``Plateau de Bure High-z Blue-Sequence Survey" (PHIBSS)}
The PHIBSS2 survey is an IRAM Plateau de Bure Interferometer \citep[PdbI;][]{guilloteau92} 4-year legacy program aimed at studying early galaxy evolution from the perspective of molecular gas reservoirs, while exploiting the capabilities of the NOrthern Extended Millimeter Array \citep[NOEMA;][]{schuster14} as they came online. Observations of $^{12}$CO\,($2-1$), $^{12}$CO\,($3-2$), and $^{12}$CO\,($6-5$) transitions took place between October 2013 and June 2017. Observation times per target range from 0.6 to 30.3 hours, with a total of $\sim 1,100$ hours of 6-antenna equivalent on source integration time, and were mostly taken in C configuration to ensure that the galaxies are not spatially resolved \citep[see][for more details on the data reduction process]{freundlich19}. Given the integration times and configurations, the synthesized beams range from 1\arcsec\ to 5\arcsec. At the redshifts targeted by PHIBSS, the typical scales are $6-8.5$ kpc per arcsec.

The survey consists of 110 individual observations of main sequence galaxies, exploring the CO\,($2-1$), CO\,($3-2$), and CO\,($6-5$) line emission, covering a total area of $\sim 130$ arcmin$^{2}$ and sampling a total co-moving volume of $\sim 200000$ Mpc$^{3}$ (see Table \ref{tab:volumes}).

\subsection{Ancillary Data}
We use the 3D-HST/CANDELS survey catalogs \citep{brammer12,skelton14,momcheva16} for the COSMOS, GOODS-N, and EGS/AEGIS fields to search for optical counterparts. We present cutouts from the HST Advanced Camera for Surveys (ACS) for filter F814W for each field (where possible) corresponding to the PHIBSS2 observations in Appendix \ref{app:spec} (see \S\ref{sec:search}). For targets lying outside the area covered by the HST ACS optical or WFC3 near-IR mosaics, we show cutouts of Spitzer IRAC 3.6 $\mu$m images.

\section{Methods} \label{sec:method}
\subsection{Line Search}
The goal of the line search is to systematically select candidate sources from noisy data, and assess their significance in terms of their corresponding signal-to-noise ratio (SNR). For our sample of observations, we expect sources to be unresolved and to mostly have FWHM in the range of $\sim 50 - 500$ km\,s$^{-1}$, and at most $\sim 1000$ km\,s$^{-1}$. \citet{rubin85} shows massive galaxies have maximum rotation velocities that span from $\sim 100 - 400$ km s$^{-1}$, while small irregular galaxies have minimum rotation speeds of $\sim 50 - 100$ km s$^{-1}$, and \citet{carilli13} show that hyper-starburst quasar hosts and sub-millimeter galaxies can have line widths up to $\sim 1000$ km\,s$^{-1}$. 

Our line search method is a 1D matched filter technique where we select a Hanning kernel as our template. We Hanning-smooth and decimate each observation five times, where each iteration of the smoothing increases the width of a channel by a factor of two while removing one every two channels (that is, decimating the highly correlated channels). This creates cubes with velocity resolutions spanning from $\sim 7$ to  $\sim 1000$ km\,s$^{-1}$, depending on the observation (the original data cubes have channel widths ranging from 7 to 88 km\,s$^{-1}$). The purpose of this matching is to maximize the signal-to-noise for signals of a given line-width. Hence our choice of smoothing allows us to attempt to match the velocity resolution of the data cube with that of the potential sources in the data. Each data cube generates five additional smoothed cubes corresponding to the different velocity resolution templates. 

\begin{figure}[!ht]
    \centering
    \includegraphics[width=\columnwidth]{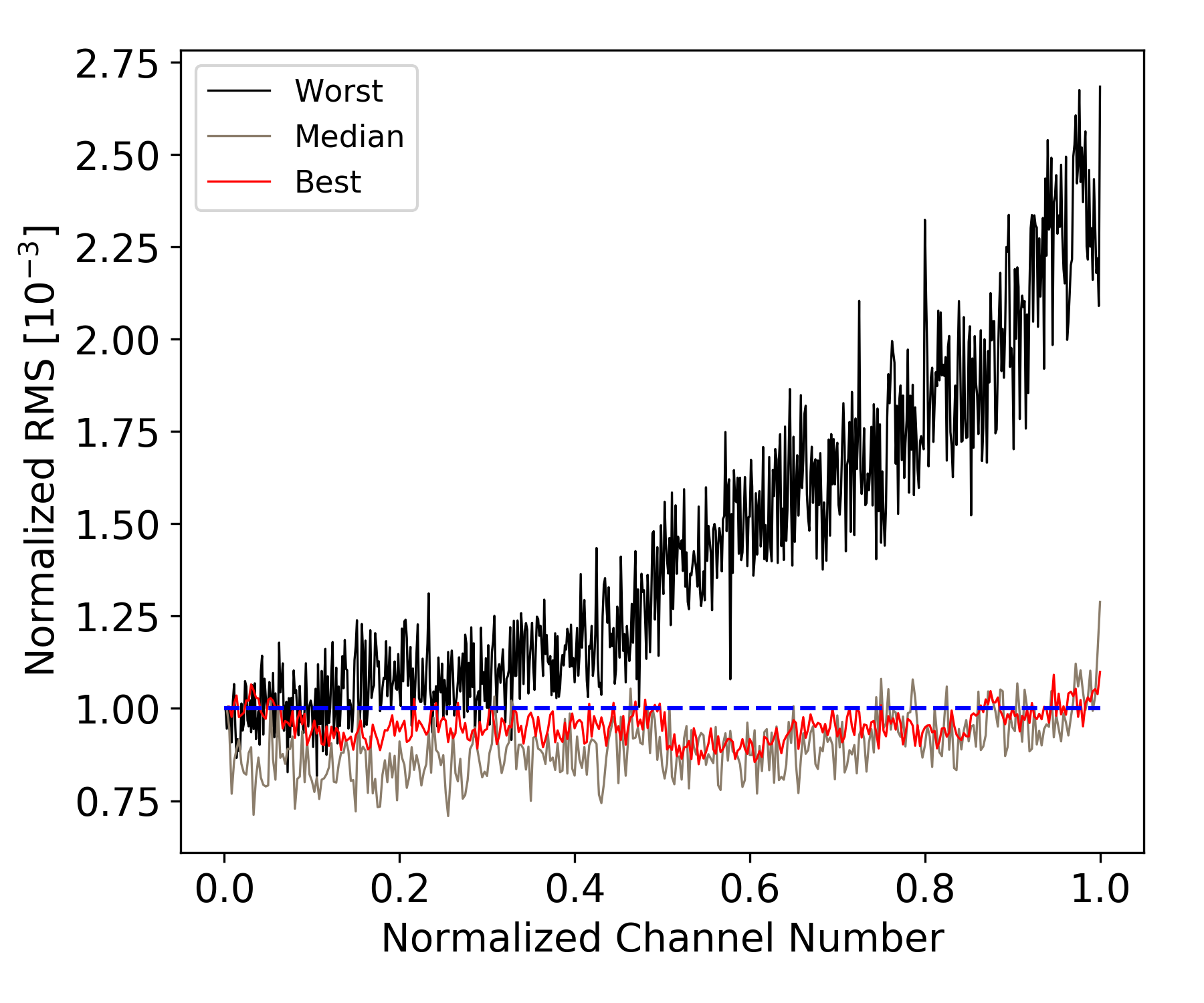}
    \caption{RMS as a function of channel number (both normalized to unity), showing a typical best case scenario (red line) where the RMS is approximately flat across all channels, a typical worst case scenario (black line) where the RMS varies quite significantly across the channels, and a median case (beige line). This illustrates the need to properly model the RMS variations across the passband in order to correctly estimate the SNR of every pixel. We do this by modeling the RMS variations as a function of frequency with a seventh order polynomial for each data cube.}
    \label{fig:RMS}
\end{figure}

For each of these cubes (original and smoothed), we compute a significance (SNR) map by taking the peak value at each pixel along the spectral axis and dividing it by the RMS (taken to be the standard deviation) of the spectral channel. The RMS as a function of channel number (frequency) is usually fairly flat, but occasionally it can vary quite substantially across the passband. We illustrate this in Figure \ref{fig:RMS}, where we show three examples of how the noise varies across channels in three different data cubes. For the purpose of comparing the RMS channel variations in different data cubes, we normalize the axes. The channel RMS values vary from one data cube to another, therefore we normalize the y-axis to $1 \times 10^{-3}$. The number of channels in the three data cubes that we compare here also varies, thus we also normalize the x-axis to unity (the bandwidth is about 3 GHz). We plot a typical best case scenario in red, a typical worst case scenario in black, and a median example in beige, while the dashed blue line serves as a reference point for a straight horizontal line. It is therefore important to properly account for this when calculating the SNR in order to not over or underestimate the SNR of a given pixel. To characterize this variation, we fit the distribution of the channel RMS as a function of channel number for each data cube with a polynomial, in order to have a smooth representation of the large-scale noise variation to properly calculate SNR. We then divide each peak pixel by the corresponding channel RMS from our fit. The order of the polynomial chosen to obtain our smooth representation of the noise is not particularly important, and a value of seven was found to produce very reasonable results.

\begin{figure}
    \centering
    \includegraphics[width=\columnwidth]{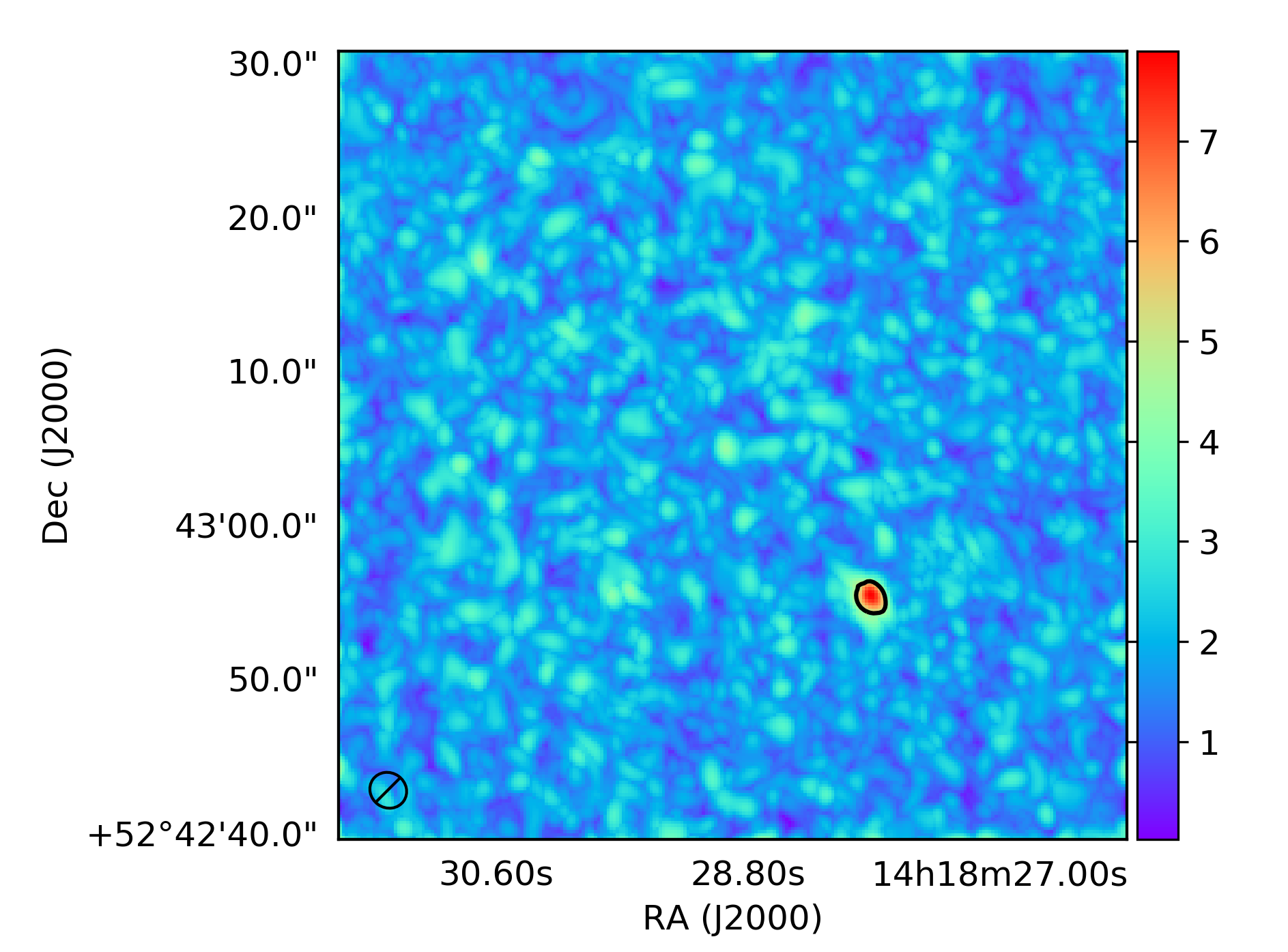}
    \caption{Hanning-smoothed SNR map for the eg016 data cube, at a velocity resolution of 352 km\,s$^{-1}$. The black contour corresponds to the SNR level of the largest negative peak in this cube, which is our detection threshold and in this case corresponds to a SNR of 4.93. A single source appears in this map with SNR above the detection threshold we impose (see eg016-1 in Table \ref{tab:props} for physical properties). The central targeted source in the eg016 data cube has a SNR of 3.1 \citep[see Table 3 in][]{freundlich19}, which is below our detection threshold and is therefore not visible in this SNR map.}
    \label{fig:snr_map}
\end{figure}

In order to obtain the distribution of the noise, we repeat this process for the negative peaks, that is dividing the largest negative peak at each pixel by the channel RMS, thus creating SNR maps of ``negative emission''. Positive emission corresponds to real astrophysical sources as well as noise peaks, while ``negative emission'' corresponds only to noise. The most significant negative peak therefore provides an estimate of what is the boundary between likely noise and likely signal. From these SNR maps, we build a list of candidate sources by selecting pixels with a positive SNR value that is greater than the absolute value of the largest negative peak SNR. We save a list of all pixels that satisfy this condition, sort it by decreasing SNR, and filter out all pixels that lie within one beam of the highest SNR pixels to arrive at a list of independent possible sources. We perform this search and filtering on all smoothed cubes and then combine the lists into one list, where we filter out candidate sources that satisfy our detection threshold in multiple cubes for a given field (original and/or smoothed cubes), but with lower SNR. This leaves us with a final list of candidate sources for each field in our sample, where the position of each source corresponds to the position of the most significant pixel for the velocity smoothing parameter that provides the highest SNR. Figure \ref{fig:snr_map} shows an example SNR map for one of our fields (eg016; see Table \ref{tab:props}) at a velocity resolution of 352 km\,s$^{-1}$, with the black contour showing the threshold of the most significant negative peak in that cube, our chosen boundary between ``likely noise'' and ``likely signal''. In what follows we estimate the probability of this candidate source being real.

\subsubsection{False Positives}
The purpose of the false positive analysis is to assign to each candidate source a probability of it being a real astrophysical source, which we will call reliability (also called fidelity or purity). To address this question, we use the statistics of the negative emission, which consists of only noise, to determine the likelihood that noise could produce a SNR as large as that of each candidate source. 

In order to estimate this we would ideally consider the statistics of independent points in the map. In our significance maps, in principle all pixels within one beam of a strong emission pixel will be correlated. To remove from our distribution of peak SNR values pixels that are correlated, we perform a ``cleaning'' of the map. We do that by taking the most significant value in a given map, subtracting a beam-like Gaussian from that pixel position, and then repeating the process until no values above 3$\times$ the RMS level of the map remain, leaving us with a list of independent ``sources'' in terms of SNR. As a comparison, we do the same thing with the SNR map distribution of positive peaks.

\begin{figure}[!ht]
    \centering
    \includegraphics[width=\columnwidth]{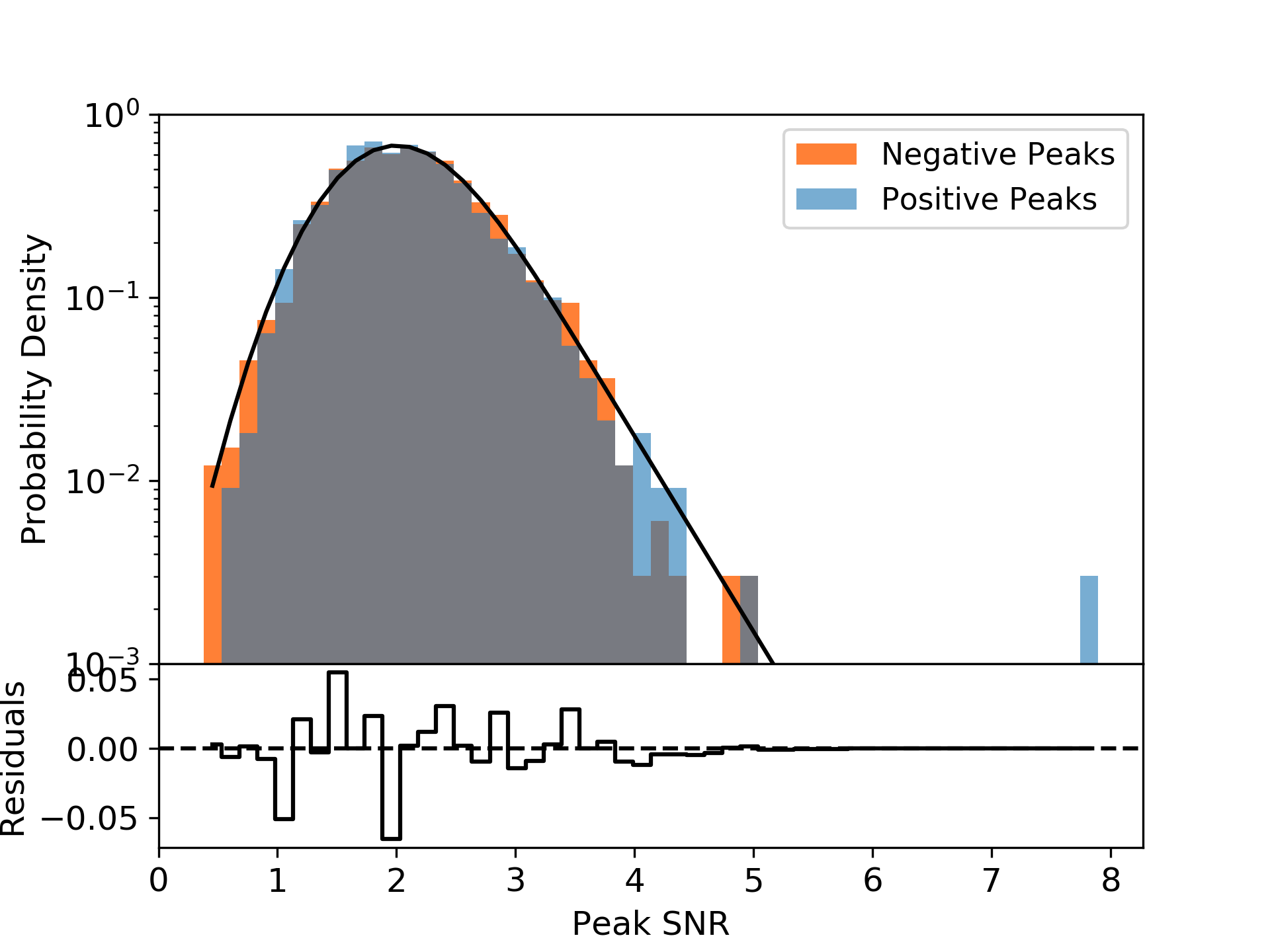}
    \caption{The top panel shows the distribution of positive (blue) and negative (orange) peak SNRs per beam of the eg016 data cube, at a velocity resolution of 352 km\,s$^{-1}$ (the gray histogram is the overlap of the blue and orange histograms). The black line is the exponentially modified Gaussian fit to the negative peaks distribution. We see one object with positive peak SNR much greater than the largest negative peak SNR (in absolute values); this corresponds to the candidate sources. The bottom panel shows the residuals from fitting with an exponentially modified Gaussian function, which we find represents that data reasonably well.}
    \label{fig:fidelity}
\end{figure}

The distributions of independent positive and negative peaks in a given map overlap very well, and are well approximated by a Gaussian with an exponential tail toward high significance (Figure \ref{fig:fidelity}). However, the tail of the distribution is the region that we are interested in characterizing because this is where the candidate sources we detect lie. To achieve this goal, we begin by normalizing the distribution of independent positive and negative peaks so that their integrals equal unity. We then treat the normalized independent negative distribution as our probability density function, which we fit with an exponentially modified Gaussian distribution of the form: 

\begin{equation}
h(x) = \frac{\lambda}{2}e^{\frac{\lambda}{2}(2\mu + \lambda\sigma^{2} - 2x)} \mathrm{erfc}\left(\frac{\mu + \lambda\sigma^{2} - x}{\sqrt{2}\sigma}\right)
\end{equation}

\noindent where $x$ corresponds to the peak SNR values of the inverted cube, $\mu$ and $\sigma$ are the mean and standard deviation of the Gaussian, $\lambda$ is the rate of the exponential, and erfc is the complementary error function which is equal to $1-$erf($x$). This function describes a Gaussian distribution with a positive skew due to an exponential component. An example of this fit is shown in Figure \ref{fig:fidelity}, where the orange histogram corresponds to the SNR distribution of the negative emission, the blue histogram is the SNR distribution of the positive emission, and the black line is the exponentially modified Gaussian; the bottom panel shows the residuals. Figure \ref{fig:fidelity} corresponds to the data cube eg016 (Table \ref{tab:props}), where one candidate source is identified as possible emission through the line search procedure described above. While the y-axis in Figure \ref{fig:fidelity} is plotted on a log-scale, the fitting procedure is done in linear space. As such the resulting parameters are not sensitive to the high-SNR ``outliers". Therefore the reliability parameters we derive from this method are robust with respect to the inclusion or exclusion of these data points.

To estimate the probability that the observed significance could be produced by noise fluctuations, we use the cumulative distribution of the exponentially modified Gaussian distribution, which has the form: 

\begin{equation}
H(x) = \Phi(u,0,\sigma) - e^{-u+v^{2}/2+\mathrm{log}(\Phi(u,v^{2},v))}
\end{equation}

\noindent where $u = \lambda(x-\mu)$ and $v = \lambda\sigma$, and $\Phi(x,\mu,\sigma)$ is the cumulative distribution function of a Gaussian distribution with mean $\mu$ and standard deviation $\sigma$.

For a given candidate source, the probability that a random fluctuation produces a source with SNR greater than or equal to that of the candidate source, (i.e., falls in the range $x$ $\in [\mathrm{SNR_{src}}$, $\infty$)), is: 

\begin{equation}
\mathrm{P}(x>\mathrm{SNR_{src}}) = 1 - H(\mathrm{SNR_{src}}).
\label{eq:prob}
\end{equation} 

For each candidate source, this gives an estimate of the probability that a given independent measurement (a beam) in the map could have a peak SNR greater than or equal to that of the candidate source itself. To assess the significance of these values, we compare them to the number of independent beams sampled by each SNR map. We do this by taking the inverse of the false positive probability we calculate from equation \ref{eq:prob} as a measure of how many random measurements it would take to observe the given candidate source SNR value once (i.e., one in every N number of measurements will have an SNR equal to or greater than what is observed for the candidate source given only noise; we call this $\mathrm{N_{expected}}$). Then the ratio of $\mathrm{N_{expected}}$ to the number of independent beams sampled ($\mathrm{N_{beams}}$) by each SNR map is the total number of measurements with a given SNR we would expect to make due to noise only. As an example, if we measure a probability of 10$^{-3}$ for a candidate source of some SNR, but then find that we sample 1000 beams in that map, then we would expect to find one such "source" in our map from just noise, so this candidate source would be considered unreliable. For very strong candidate sources this number is very small, and for weaker sources it becomes larger and can become on the order of unity. The reliability parameter ($R$) we assign to each source is one minus this ratio:

\begin{equation}
    R = 1 - \frac{\mathrm{N_{expected}}}{\mathrm{N_{beams}}}
\end{equation}

Our reliability measurements range from $0.01-1$ (i.e., $1-100$\% reliability), and we include in our sample candidate sources with $R > 5\%$ since this is the threshold adopted by \citet{riechers19}. We show these values in Table \ref{tab:props}. Note that we do not attempt to further filter our list of candidate sources by choosing a higher reliability cutoff. Note also that this definition of reliability is more conservative than the ``fidelity'' parameter employed by \citet{decarli16}, for example, as per our definition there are no candidate sources with lower flux than the absolute value of the largest negative peak in a map. There is a strong correlation between integrated flux and reliability, where fainter sources with lower SNR naturally tend to show lower $R$ (see \S\ref{sec:cen_sec}, Figure \ref{fig:gas_flux}). The derivation of the luminosity function (\S\ref{sec:lfs}) properly takes into account the statistics by weighting by reliability, and artificially inserting a high reliability cutoff would cause us to preferentially remove the contribution from fainter sources. Note also that computation of $R$ for the central sources, all known to be real, shows a large spread driven by SNR. So it is clear that real sources can have low reliability when they are faint in relation to the noise of the observation.

\subsubsection{Completeness}
To assess the completeness of our search algorithm, we perform an analysis of the chance of detecting sources we artificially inject into each data cube. The purpose of this analysis is to relate the fraction of recovered simulated sources to the line flux. Since we do not expect resolved sources in the PHIBSS2 data, we do not account for varying sources sizes. 

To simulate sources, we assume a Gaussian line profile along the spectral axis, and generate sources with five free parameters: the spatial position, the peak flux density of the line, its velocity width as FWHM, and the velocity of the peak by drawing random numbers from a uniform distribution. The x and y coordinates are limited to between 1 and 256, since the cubes are 256 $\times$ 256 pixels in size. We test the effect of source position on the completeness by simulating sources at the edges of pixels and at the centers of pixels, and find that this has a negligible impact on the completeness correction factors. The peak flux density of each artificial source ranges between the maximum value in the data and 1\% of the maximum. Because completeness is also a function of line width, we simulate sources with FWHM values ranging from $50 - 1000$ km s$^{-1}$, to reflect the range of line widths spanned by the data.

\begin{figure}
    \centering
    \includegraphics[width=\columnwidth]{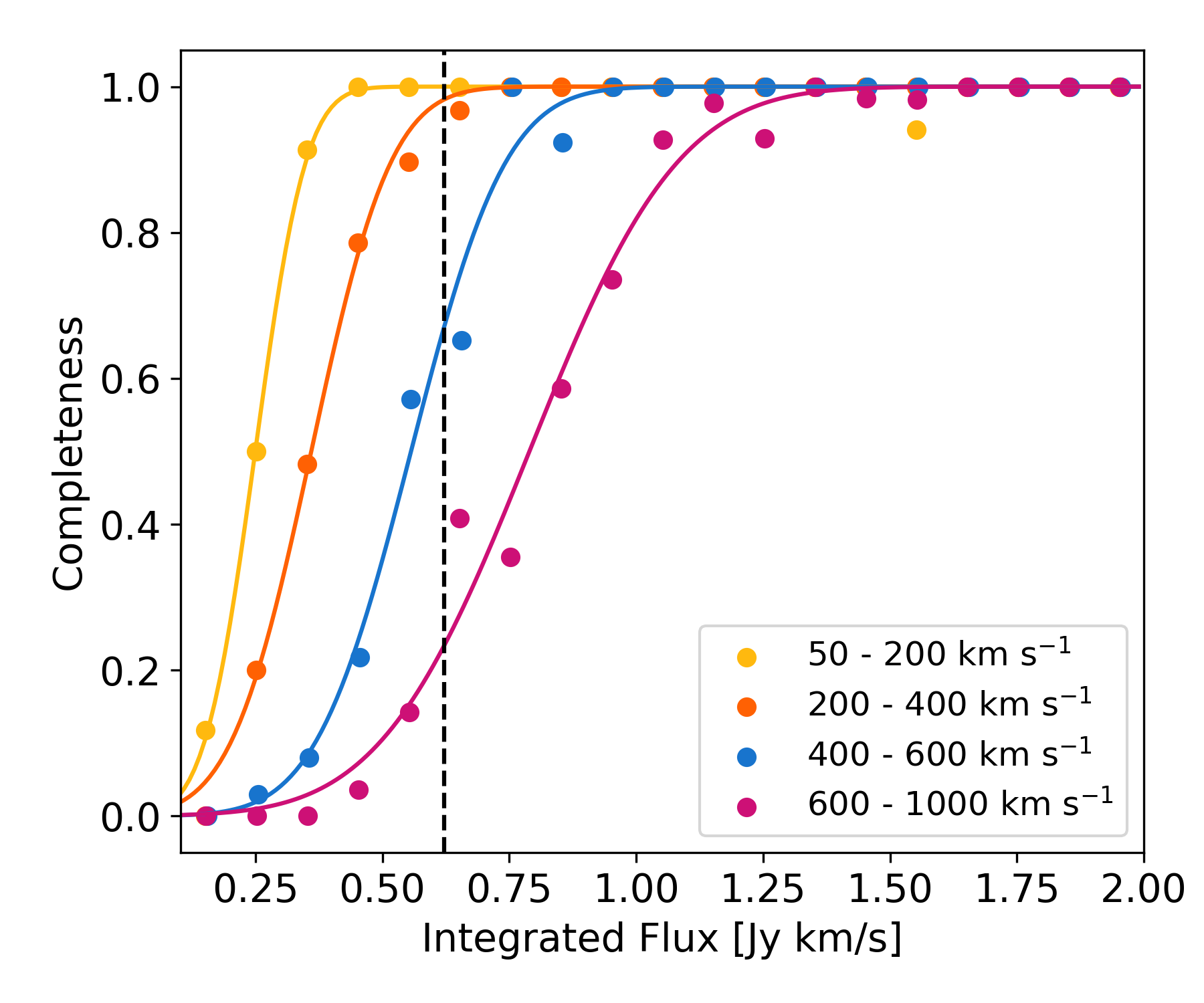}
    \caption{The fractions of recovered sources to artificial sources injected as a function of integrated flux and FWHM for the eg016 field, from our analysis of 2500 simulated sources. The colored data points correspond to the fraction of recovered sources for four velocity bins. The colored lines are fits to the four distributions using a cumulative Gaussian distribution. The vertical dashed lines correspond to the integrated flux of the candidate source. As we would expect, the recovery of sources decreases with decreasing integrated flux indicating that fainter sources are harder to detect than brighter ones. We also see that the recovery of sources at a given integrated flux decreases with increasing FWHM. This analysis allows us to correct our CO luminosity functions for the incompleteness of our search algorithm, particularly at the faint end where this becomes a larger effect (see \S\ref{sec:lfs}).}
    \label{fig:comp}
\end{figure}

We then assume that each source will be ``beam-like", so we take the flux density at each velocity channel that the source appears in to be the peak of a two-dimensional Gaussian which has the same position angle and size as the synthesized beam for each data cube. However, since power is lost in the side lobes of the full synthesized beam, we take this into account by correcting the flux of each simulated source by the ratio of power in the primary lobe of the beam and its side lobes. We generate 2500 artificial sources for each data cube in this way, add them to the data cube 5 sources at a time to avoid crowding, run the search algorithm, and check the fraction of sources recovered. Figure \ref{fig:comp} plots the fraction of recovered artificial sources as a function of integrated flux (blue circles), for the eg016 data cube. The recovered fraction is fit with a Gaussian cumulative distribution function (solid blue line). The vertical dashed black lines correspond to the integrated flux of the candidate sources for this data cube. We can see that the recovery fraction decreases with decreasing integrated flux, which is known for each simulated source. We correct for completeness on a source-by-source basis using the cumulative Gaussian distribution fit for each data cube. Given the integrated flux of each candidate source, $x$, the corresponding completion correction is

\begin{equation}
    C(x) = \frac{1}{2}\left[1 + \mathrm{erf}\left(\frac{x - \mu}{\sigma\sqrt{2}}\right)\right],
\end{equation}

\noindent where $\mu$ and $\sigma$ are the mean and standard deviation derived from our Gaussian cumulative distribution function fit for a given cube. Primary beam attenuation, the sensitivity drop off as a function of distance from the pointing center, will decrease the chances of detecting weaker sources closer to the edges of each data cube. We take this effect into account in our comoving volume calculations, which results in a smaller effective volume sampled by each cube.

\section{Results of Line Search} \label{sec:results}
\subsection{Line Properties}
We extract the spectrum of each candidate source at the position of the peak SNR pixel, given that the sources in PHIBSS2 are unresolved, in each field at all velocity resolutions, and apply a primary beam correction. These spectra are fit with a Gaussian profile using Python's \texttt{scipy.optimize.curve\_fit}. Example spectra of the three brightest candidate sources detected in the COSMOS, EGS/AEGIS, and GOODS-N fields are shown in Figure \ref{fig:sources} in Appendix \ref{app:spec}, while the remaining set of figures is available in the online journal.

The redshift of each candidate source is calculated from the central frequency of the line, assuming that the emission detected corresponds to a CO transition from CO\,($1-0$) to CO\,($6-5$). CO emission represents usually the brightest line in galaxy spectra at wavelengths between 400 and 2600 $\mu$m. Rotational transitions of CO are spaced by intervals of 115.27 GHz, so with a single transition by itself it is impossible to determine the redshift of the source. The optical counterpart search discussed in the next section allows us to, in some cases, determine which CO transitions a candidate source may correspond to, and in other cases, to constrain the range of possible CO transitions. 

The flux and full-width-half-maximum of each candidate source are calculated from the best fit standard deviation and amplitude of the Gaussian profile fit. These results are presented in Table \ref{tab:props} in Appendix \ref{app:spec}.

\subsubsection{Optical Counterparts}\label{sec:search}
The purpose of identifying counterparts (CPs) for the candidate sources is to constrain their likely redshift and CO transition, as well as properties like their stellar masses and SFRs. We search for all optical sources in the 3D-HST/CANDELS catalogs \citep{brammer12,skelton14,momcheva16} that lie within one beam FWHM radius of the peak SNR position of each candidate source we identify in PHIBSS2, while leaving the redshifts unconstrained. The objects in these catalogs have a distribution of redshifts determined from HST and ground-based spectral energy distribution (SED) fitting using the EAZY code \citep{brammer08}. To match the redshifts, we then consider all transitions from CO\,($1-0$) to CO\,($6-5$) and check which, if any, CO transitions are plausible given the posterior likelihood distributions of the redshift determination from the SED fitting. In several cases the redshifts of the optical counterparts are poorly constrained by the SED fitting, allowing a range of possible CO transitions. When grism or spectroscopic redshifts are available, we compare our redshifts to those because they are much better constrained than the photometric redshifts. For the purpose of constructing the CO luminosity functions, we assign a ``redshift probability'' to each source based on the posterior likelihood distribution. We also assign an ``association probability'' for candidate CO sources where multiple optical counterpart candidates lie within the synthesized beam of the CO data cube (which changes from cube to cube). This ``association probability'' is defined as $P_{a} = 1-(\Delta r^{2}/\theta)$, where  $\Delta r$ is the projected angular separation between the CO source and potential counterpart, and $\theta$ is the synthesized beam area. In this way, optical sources that lie outside of the synthesized beam area are assigned an association probability of zero, and the probability of association increases as the projected angular separation decreases.

From this spatial matching, and CO transition/redshift association, we find that $\sim$ 64\% (43 out of 67) of source candidates in our catalog have at least a tentative optical counterpart. The lack of an optical counterpart in the 3D-HST/CANDELS catalog (rest frame optical/UV counterparts) could imply that the candidate source is spurious, though based on our reliability calculations, we would not expect more than 25\% of sources to be spurious. Thus, this could also be physically caused by heavy extinction associated with the molecular gas (in which case there may be infrared counterparts). \citet{whitaker17} investigate the relation between dust obscured star formation and stellar mass as a function of redshift ($z = 0-2.5$). They find that for log($M/M_{\odot}$) $> 10.5$, more than 90\% of star formation is obscured by dust at all redshifts, and that at $z>1$, there is a tail of heavily obscured low-mass star-forming galaxies. This highlights the importance of infrared data, and future work may involve carrying out a systematic infrared counterpart search beyond existing catalogs (e.g., \textit{Spitzer} IRAC 3.6 and 4.5 $\mu$m). 

The results of our search are presented in the middle panel of the figures in Appendix \ref{app:spec}. These are for the most part HST ACS F814W images where the red crosses mark the positions of the candidate optical counterparts for candidate CO sources where one could be tentatively identified. For candidate sources where no ACS optical and/or WFC3 near-IR data was available, we present \textit{Spitzer} IRAC 3.6 $\mu$m images. In the left panel, the redshifts reported correspond to the CO transition that most closely matches the ``best'' redshift reported in the 3D-HST/CANDELS catalogs. In the computation of the CO luminosity functions, we however use the range of possible CO transitions/redshifts allowed by the potential counterparts to derive CO luminosities, weighted by their respective probabilities (see \S\ref{sec:lfs} for details). Finally, these results are also summarized in Table \ref{tab:cntrprts} where we give the right ascension, declination, and ``best'' redshift reported in the 3D-HST/CANDELS catalogs of each optical counterpart. We also provide the CO based redshifts for the range of possible CO transitions as determined from the EAZY SED fitting posterior likelihood distributions. Finally, we provide the angular separation between the candidate source and potential optical counterpart, with a probability of association in cases where more than one possible counterpart exists within the synthesized beam.

\begin{figure*}
    \centering
    \includegraphics[width=\textwidth]{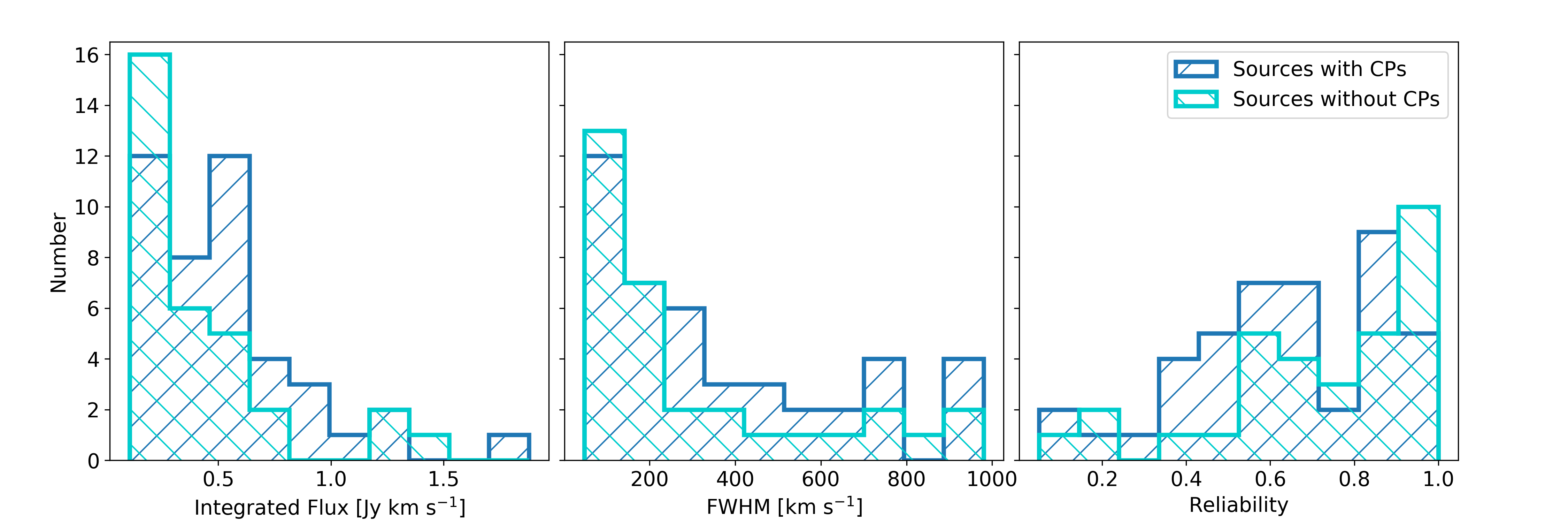}
    \caption{Comparing the properties of the candidate sources with optical counterparts (dark blue, right-hatched histogram) to those without (cyan, left-hatched histogram). \textit{Left}: comparing the integrated flux, \textit{Middle}: comparing the line width (FWHM), \textit{Right}: comparing the reliability. The K-S test results indicate that the distributions are similar in terms of all three properties.}
    \label{fig:sec_comp}
\end{figure*}

In Figure \ref{fig:sec_comp} we compare the integrated flux, line width, and reliability of candidate sources with potential optical counterparts (dark blue, right-hatched histogram), and those without (lighter blue, left-hatched histogram). In all three cases, both populations of candidate sources span the same parameter space. Both populations contain many fainter objects and fewer bright objects, so while some of those may be spurious detections, the reliability distribution shows that there are several high-reliability objects with no optical counterpart identified. In terms of line width, both populations span essentially the same range of line widths probed. 

To quantify this we perform the Kolmogorov-Smirnov test and find a K-S statistic of D$_{n,m}=$ 0.21, 0.25, and 0.36 for the integrated flux, FWHM, and reliability distributions respectively (where $n$ and $m$ are the lengths of the two samples). The K-S statistic is the maximum distance between the cumulative distributions of the two compared populations, so a small enough K-S statistic indicates that the hypothesis that two samples are drawn from the same distribution cannot be rejected. Specifically, the two samples can be said to come from different distributions at a confidence level $\alpha$ if 
\begin{equation}
    \mathrm{D}_{n,m} > c(\alpha) \sqrt{\frac{n+m}{nm}}
    \label{eq:KS}
\end{equation}

\noindent where 

\begin{equation}
    c(\alpha) = \sqrt{-\frac{1}{2}\mathrm{ln}\alpha}.
\end{equation}

From the K-S statistics for these three distributions, we find that the hypothesis that both samples are drawn from the same distribution can be rejected at the 74.4\%, 86.1\%, and 98.1\% confidence level for the integrated flux, FWHM, and reliability respectively. These confidence levels are usually not considered significant enough to reject the hypothesis. We conclude from this that we lack evidence to say that the two populations are different and note only that the candidate sources without counterparts tend to be fainter and consequently less reliable that the candidate sources with counterparts.

For candidate sources where we identify possible counterparts (and hence for which we have a redshift $z$), we compare the molecular gas mass from the CO luminosity to the molecular gas inferred from the potential counterpart SFR, using the depletion time scale scaling relation of \citet{tacconi18}:

\begin{equation}
    \mathrm{log}(t_{dep}) = A_{t} + B_{t}\,\mathrm{log}\,(1+z) + C_{t}\,\mathrm{log}\,(\delta \mathrm{MS})
\end{equation}

where $A_{t} = 0.09$, $B_{t} = -0.62$, $C_{t} = -0.44$ \citep[for details see][]{tacconi18}, and $\delta$MS is the offset from the main sequence of a source. Using the redshift, and main sequence offset of the potential counterpart, we calculate their depletion timescales and then infer the molecular gas mass based on their SFR (since $t_{dep} = M_{gas}/\mathrm{SFR}$).

\begin{figure}
    \centering
    \includegraphics[width=\columnwidth]{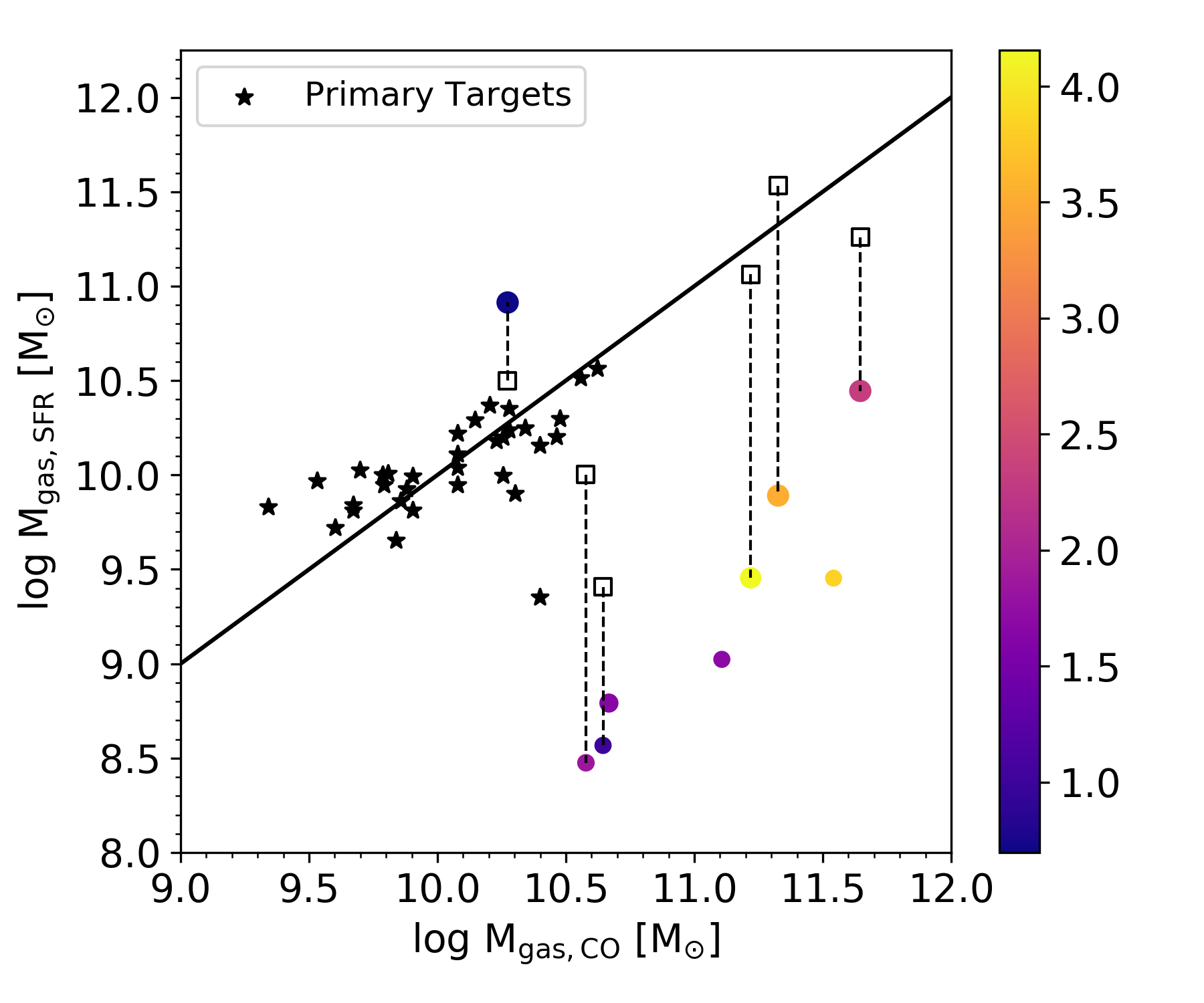}
    \caption{Comparison of the molecular gas mass measured from the candidate source CO luminosities to the molecular gas mass inferred from the potential optical counterpart SFR and the depletion timescale scaling relation of \citet{tacconi18}. The size of the colored points is scaled according to the product of reliability and association probability of the detection, and they are colored according to redshift. The diagonal black solid line is the one-to-one relation. All colored symbols correspond to SFR measurements from SED modeling of optical to 8 $\mu$m photometry; the black square symbols show the effect of including longer wavelength photometry (24 or 160 $\mu$m) on the SFR calculation for the sources where that is available.}
    \label{fig:gas_comp}
\end{figure}

We plot this comparison in Figure \ref{fig:gas_comp}, omitting candidate sources with multiple possible counterparts identified within one synthesized beam and sources where the product of the CO source reliability ($R$) times the counterpart probability of association ($P_a$) is less than 50\%, as we consider these sources and/or counterparts not highly reliable. The size of the data points is scaled according to the product of the reliability and the probability of association (higher $R\times P_a$ correspond to larger symbols), and colored according to redshift. The black solid line is the one-to-one relation and the black stars are primary PHIBSS2 targets plotted as a comparison. These all lie on the one-to-one line, except for one target, which has a large offset from the main sequence of star formation ($\log \delta{\rm MS} = +2.41$, corresponding to a target over the main sequence) and therefore a very short depletion timescale. In contrast, the majority of potential counterparts lie systematically below the one-to-one relation, which would imply molecular gas reservoirs larger than would be inferred from the measured star formation. 

The SFRs reported in the 3D-HST/CANDELS catalog for these objects are derived from SED modeling. We have used the catalog by \citet{momcheva16}, but with SFR values recomputed according to the Herschel-calibrated ladder of indicators in \citet{wuyts11} \citep[see also][and references therein]{tacconi18}. The SEDs for all objects contain optical to 8\,$\mu$m photometry, and some objects have photometry at longer wavelengths. At the redshifts of these objects, 8 $\mu$m corresponds to rest-frame wavelengths of $\lambda\sim1.5-4$ $\mu$m. For six data points the photometry also includes 24 $\mu$m to 160 $\mu$m measurements. These are indicated by vertical dashed lines, which join the SFR obtained from fitting the $\lambda\leq 8$\,$\mu$m photometry to the SFR computed including the longer wavelengths (which corresponds to the square symbols in Figure \ref{fig:gas_comp}). When the SED modeling includes only the shorter wavelengths, it results in SFRs one to two orders of magnitude lower than is estimated when longer wavelength data are included. The agreement between molecular masses estimated from the optical counterpart star formation activity, and those directly measured in the PHIBSS2 observations, is very good when the SFR estimate includes $\lambda\geq24$\,$\mu$m information.

Our identification process naturally selects objects that are bright in CO, and indeed they all have very large molecular masses as inferred from their flux. Therefore they are likely dust-rich, and their star formation activity is highly extincted. It appears likely that the dust-obscured component of star formation is not properly accounted for when the longest rest-frame wavelength included in the SED is $\lambda\sim1.5-4$ $\mu$m. We believe this is the main cause for the majority of the large discrepancies between the two estimates of molecular gas mass. It is also possible, particularly for sources with low reliability or probability of association, that some of them are not real or that some counterparts are misidentified. The agreement between the CO luminosity function we derive from these data (\S\ref{sec:lfs}) and other measurements in the literature, however, suggests that this is not the case for the majority of our objects.

\subsection{Comparing Serendipitous Detections to Central Sources}\label{sec:cen_sec}
The goal of PHIBSS2 is to study galaxy evolution from the perspective of molecular gas reservoirs. Surveys such as PHIBSS2 that target specific galaxies selected based on their stellar mass, SFR, and availability of ancillary data have complex selection functions. The blind search we have performed here, and our catalog of serendipitous detections provide a sample of objects that are mostly free of selection biases, other than the selection function imposed by the redshift ranges surveyed in any given observation and the flux which makes brighter objects easier to detect. We can therefore compare these two samples to get an idea of their respective biases.

\begin{figure*}[!ht]
\begin{center}
\includegraphics[width = \textwidth]{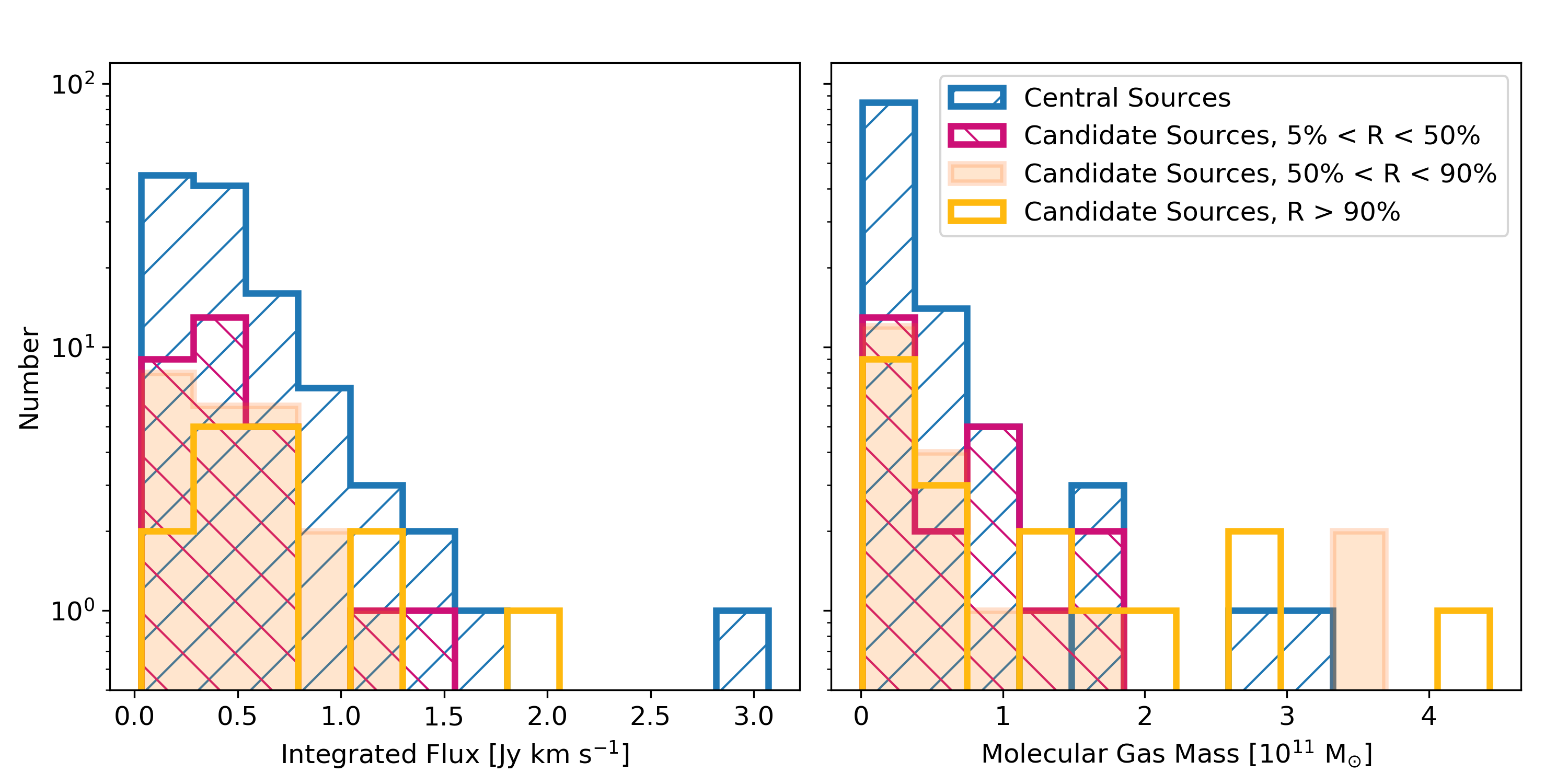}
\caption{\textit{Left}: Comparison of the integrated flux measurements of the central galaxies that were specifically targeted by PHIBSS2 (blue hatched histograms) to the additional serendipitous CO detections. The candidate sources are divided according to their likelihood parameter. The candidate sources generally seem to follow a similar distribution of fluxes as the targeted central sources. A K-S test reveals that at the 48.2\%, 94\%, and 96.6\% confidence level, the candidate source distributions do not come from the same distribution as the central sources (for the $5-50\%$, $50-90\%$, and $>90\%$ reliability ranges). These weak rejections suggest that the samples are representative of the same parent population of objects. \textit{Right}: The same as the left panel, but now comparing the molecular gas masses. The highest reliability objects tend to have the higher molecular gas masses.}
\label{fig:gas_flux}
\end{center}
\end{figure*}

In the left panel of Figure \ref{fig:gas_flux}, we compare the integrated fluxes of all 67 candidate sources to those of the central sources targeted by PHIBSS2. In the right panel of Figure \ref{fig:gas_flux}, we compare the molecular gas masses of the candidate sources with tentative optical counterpart identifications to that of the central sources. The central sources are plotted as the blue hatched histogram and the candidate sources are separated into histograms corresponding to likelihood levels: the hatched magenta histogram corresponds to sources with reliabilities between 5 and 50\%, the orange filled histogram corresponds to sources with reliabilities between 50 and 90\%, and the yellow histogram corresponds to sources with reliabilities greater than 90\%. 

We see in the flux comparison that the sample of central sources and the sample of secondary candidate detections seem to generally probe objects with similar properties. To quantify this observation and determine if candidate sources with lower reliabilities have systematically different integrated flux properties, we perform a K-S test. We find that D$_{n,m}=$ 0.13, 0.19, 0.34, which results in rejecting at the 54.0\%, 73.3\%, and 95.4\% confidence level the hypothesis that the candidate source distributions come from the same distribution of central sources for the $5-50\%$, $50-90\%$, and $>90\%$ reliability ranges respectively. These are not strong rejections, suggesting that regardless of the reliability, the candidate secondary sources have properties that are very similar to those of the central targeted sources. In terms of molecular mass, our higher reliability candidate sources seem to correspond to slightly more massive objects not well represented in the original PHIBSS2 sample, selected to represent the main sequence at the redshifts of interest. In both panels we see that the fainter/less massive candidate sources tend to have lower reliabilities than the brighter/more massive objects. This is not surprising, since these candidates will have lower SNRs.

\begin{figure*}
    \centering
    \includegraphics[width=\textwidth]{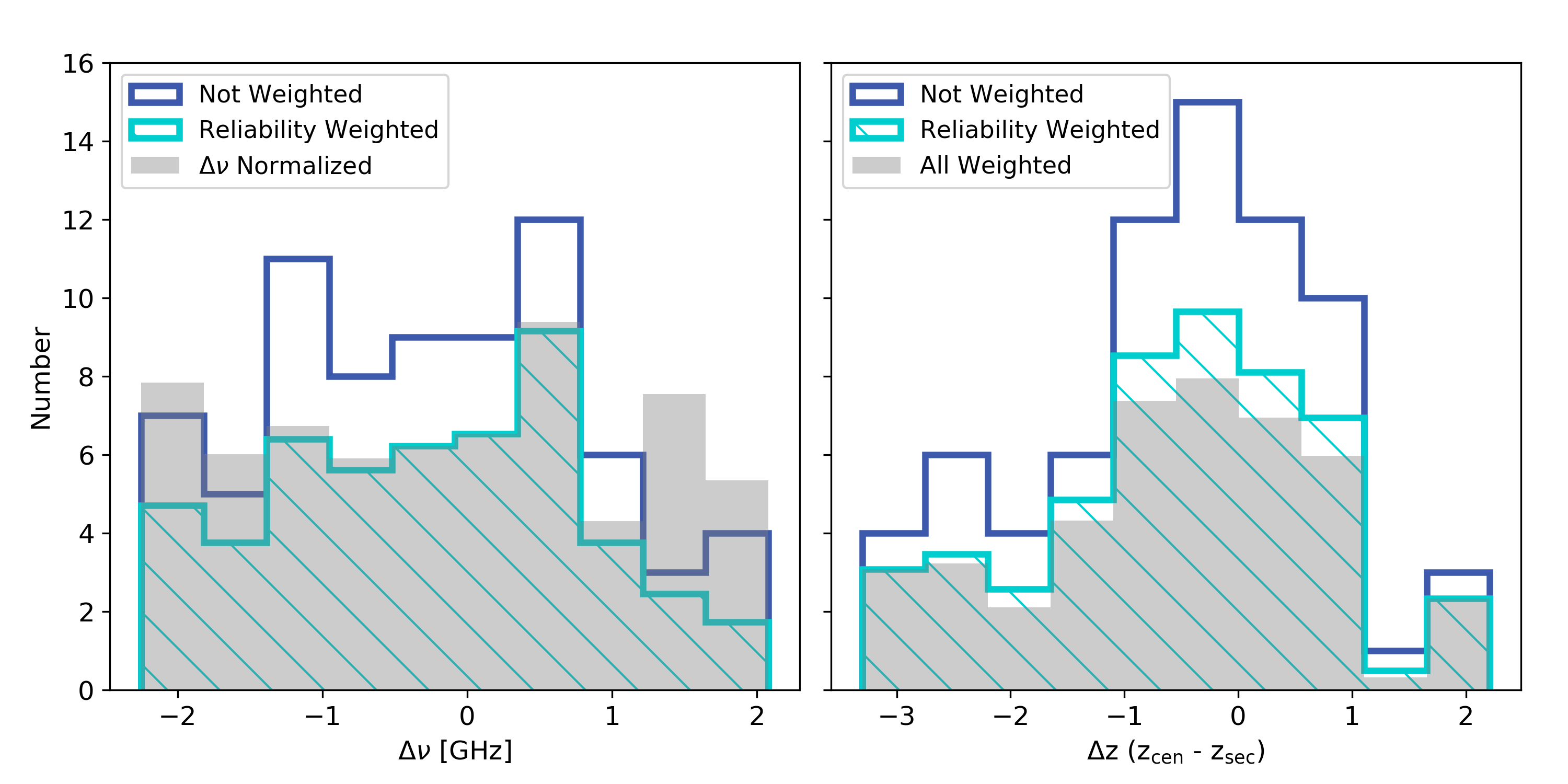}
    \caption{\textit{Left}: Difference between the central frequency of each candidate source and the reference frequency of the observation, $\Delta\nu$. The dark blue empty histogram is unweighted by reliability, while the cyan left hatched histogram is weighted by reliability. The grey shaded histogram is the reliability weighted distribution normalized to the number of data cubes that cover a large enough frequency range to reach a given $\Delta\nu$ value. For randomly distributed objects, we would expect a flat distribution and this is what we observe. \textit{Right}: Difference between the redshift of the central source and candidate source (only for cases where a tentative optical counterpart is identified), $\Delta z$. The dark blue empty histogram is the unweighted data, the cyan hatched histogram is weighted by reliability, and the grey shaded histogram is weighted by reliability, the probability of association, and the redshift probability. There is a tendency here for objects to cluster around $\Delta z \pm 1$, however this is too large of a redshift separation to form physical associations. We conclude the candidate sources we detect are not biased by clustering around the central source.}
    \label{fig:freq_z}
\end{figure*}

We observe across our sample of candidate detections and tentative optical counterpart identifications that some candidate sources lie at redshifts similar to that of the central target source. This raises the question of whether constructing a CO luminosity function from data targeted at particular objects introduces biases due to possible clustering of sources around the targeted object. To evaluate whether this is the case, we compare in Figure \ref{fig:freq_z} the difference between the frequency of each candidate source and the frequency of the central source in each data cube($\Delta\nu$; left panel). We also show the difference between the redshift of the central source and candidate source, for candidates with identified counterparts ($\Delta z$; right panel). In both cases we also compare the distribution of $\Delta\nu$ and $\Delta z$ when weighting the data by reliability, probability of association, and redshift probability. 

The left panel of Figure \ref{fig:freq_z} shows in both the unweighted and weighted cases that the candidate sources are approximately uniformly distributed in $\Delta\nu$ with a slight decrease for $\Delta\nu \gtrsim 1$ and a bit of a central bump for completely unweighted sources. However, the data cubes do not all cover the same frequency range and therefore the chance of a source to show at a particular $\Delta\nu$ has to be weighted accordingly. To account for this, we normalize the reliability weighted histogram by the number of data cubes that span the different possible $\Delta\nu$ ranges. This is shown in the gray histogram, where we see that the recovered distribution is very consistent with a uniform distribution across the spectral range. This shows that our secondary detections are uniformly distributed in $\Delta\nu$, and therefore there is no signature of a bias introduced by clustering around the targeted central sources. In physical terms, $\Delta\nu=1$~GHz for a source at $z\sim1.5$ in a $\lambda\simeq3$\,mm observation represents a physical velocity difference of over 3,000 km\,s$^{-1}$, larger than the central velocity dispersion of a massive galaxy cluster like Coma \citep[$\sigma_V\sim1200$ km\,s$^{-1}$,][]{kent82}. Therefore we would expect a relatively narrow peak in the corrected histogram if most sources were physically related to the central source, independent of our ability to identify counterparts.

The right panel of Figure \ref{fig:freq_z} shows the distribution in $\Delta z$ for only those candidate sources for which we find tentative optical counterparts. The unweighted case shows a wide peak in the distribution of objects at $z \pm 1$ from the central sources. When weighting by reliability, probability of association, and redshift probability this peak is significantly smoothed but still present. The existence of a broad peak is to be expected: most of our observations target the $2-1$ and $3-2$ CO transitions at $z\sim 1-2$, and the most likely bright transitions for field objects will be $2-1$ to $4-3$, which would place them in the $\Delta z\sim\pm\,1$ range for most observations. Note also that if this were an indication of true physical clustering we would expect the peak to be much narrower, $\Delta z\lesssim\pm\,0.1$.

\subsection{CO Luminosity Functions}\label{sec:lfs}
We construct the CO luminosity functions using equation \ref{eq:co_lf}:

\begin{equation}
\Phi(\mathrm{log}L_{i}\mathrm{)} = \frac{1}{V} \sum_{j=1}^{N_{i}} \frac{R_{j}}{C_{j}} \, P_{a,j} \, P_{z,j}\,.
\label{eq:co_lf}
\end{equation}

Here $N_{i}$ is the number of galaxies that fall within the luminosity bin $i$ defined by log $L_{i} - 0.25$ and log $L_{i} + 0.25$ (log $L_{i} - 0.5$ and log $L_{i} + 0.5$ for cases where we only have a small number of sources), V is the total volume of the Universe that is sampled by a given transition across all of our data cubes, $\mathbf{R_{j}}$ is the reliability of the $j^{th}$ line and $C_{j}$ is its completeness, $P_{a,j}$ is the probability that the  candidate source is associated with a particular optical counterpart, and $P_{z,j}$ is the probability that a given candidate optical counterpart corresponds to a particular CO transition (and hence redshift). Each CO line is down-weighted by its likelihood probability calculated in \S\ref{sec:method}, probability of association, and redshift probability, and then up-scaled by its completeness fraction. The CO luminosities are calculated from equation 3 of \citet{solomon97}:

\begin{equation}
L'_{CO} = 3.25 \times 10^{7} \frac{S_{\mathrm{CO}} \Delta V D_{L}^{2}}{\nu_{obs}^{2} (1+z)^{3}} \mathrm{\quad [K\; km\; s^{-1}\; pc^{2}]}
\label{eq:co_lum}
\end{equation}

\noindent where $S_{CO} \Delta V$ is the integrated flux density in units of Jansky kilometers per second, $D_{L}$ is the luminosity distance of the source in megaparsecs, $\nu_{obs}$ is the observed frequency of the line in GHz, and $z$ is its redshift. The volume of the Universe that is sampled by a given PHIBSS2 data cube is calculated as a three-dimensional slab of space defined by the field-of-view of the given observation, and frequency range that is observable by the instrument, for each CO transition we consider in our counterpart search. These values are summarized in Table \ref{tab:volumes}. 

\begin{deluxetable}{lccccc}
\tablecaption{Comoving Volume Sampled by each CO Transition. \textbf{The sensitivity drop off due to the primary beam is accounted for in the volume calculations.} \label{tab:volumes}}
\tablewidth{700pt}
\tabletypesize{\scriptsize}
\tablehead{
	\colhead{Transition} 			&
	\colhead{$\nu_{\mathrm{rest}}$} &
	\colhead{$z_{\mathrm{min}}$}    &
	\colhead{$z_{\mathrm{max}}$}    &
	\colhead{Volume} 				&
	\colhead{CV}       \\ 
	\colhead{}				 		&
	\colhead{[GHz]}				 	&
	\colhead{}				 		&
	\colhead{}				 		&
	\colhead{[Mpc$^{3}$]}			&
	\colhead{[\%]}                  
} 
\startdata
\cutinhead{PHIBSS2}
CO\,($2-1$) & 230.538 & 0.017 & 1.562 & 11250 & 18.2 \\
CO\,($3-2$) & 345.538 & 0.492 & 2.843 & 26136 & 15.9 \\
CO\,($4-3$) & 461.041 & 0.989 & 4.124 & 36144 & 14.9 \\
CO\,($5-4$) & 576.268 & 1.486 & 5.405 & 42380 & 13.3 \\
CO\,($6-5$) & 691.473 & 1.983 & 6.685 & 46288 & 15.6 \\
\cutinhead{COLDz COSMOS}
CO\,($1-0$) & 115.271 & 1.953 & 2.723 & 20189 & 36.9 \\
CO\,($2-1$) & 230.538 & 4.906 & 6.445 & 30398 & 37.8 \\
\cutinhead{COLDz GOODS-N}
CO\,($1-0$) & 115.271 & 2.032 & 2.847 & 131042 & 25.5 \\
CO\,($2-1$) & 230.538 & 5.064 & 6.695 & 193286 & 25.6 \\
\cutinhead{ASPECS LP}
CO\,($1-0$) & 115.271 & 0.003 & 0.369 & 338 & 59.4 \\
CO\,($2-1$) & 230.538 & 1.006 & 1.738 & 8198 & 36.9 \\
CO\,($3-2$) & 345.538 & 2.008 & 3.107 & 14931 & 35.0 \\
CO\,($4-3$) & 461.041 & 3.011 & 4.475 & 18242 & 35.2 \\
\enddata
\end{deluxetable}

We exclude from our CO luminosity functions all central sources since these were targeted objects and are therefore not the result of our blind search. We also exclude objects with no optical counterpart identification, because we have no information on their corresponding redshift or CO transition. We note that because we choose to not include sources for which we identify no counterpart, that care should be taken when comparing our results to previous blind survey results in subsequent sections and figures. The constraints we derive should be considered lower limits since the catalogs we draw counterparts from may be incomplete.

\subsubsection{PHIBSS2 CO Luminosity Functions}
\begin{figure*}
    \centering
    \includegraphics[width=\textwidth]{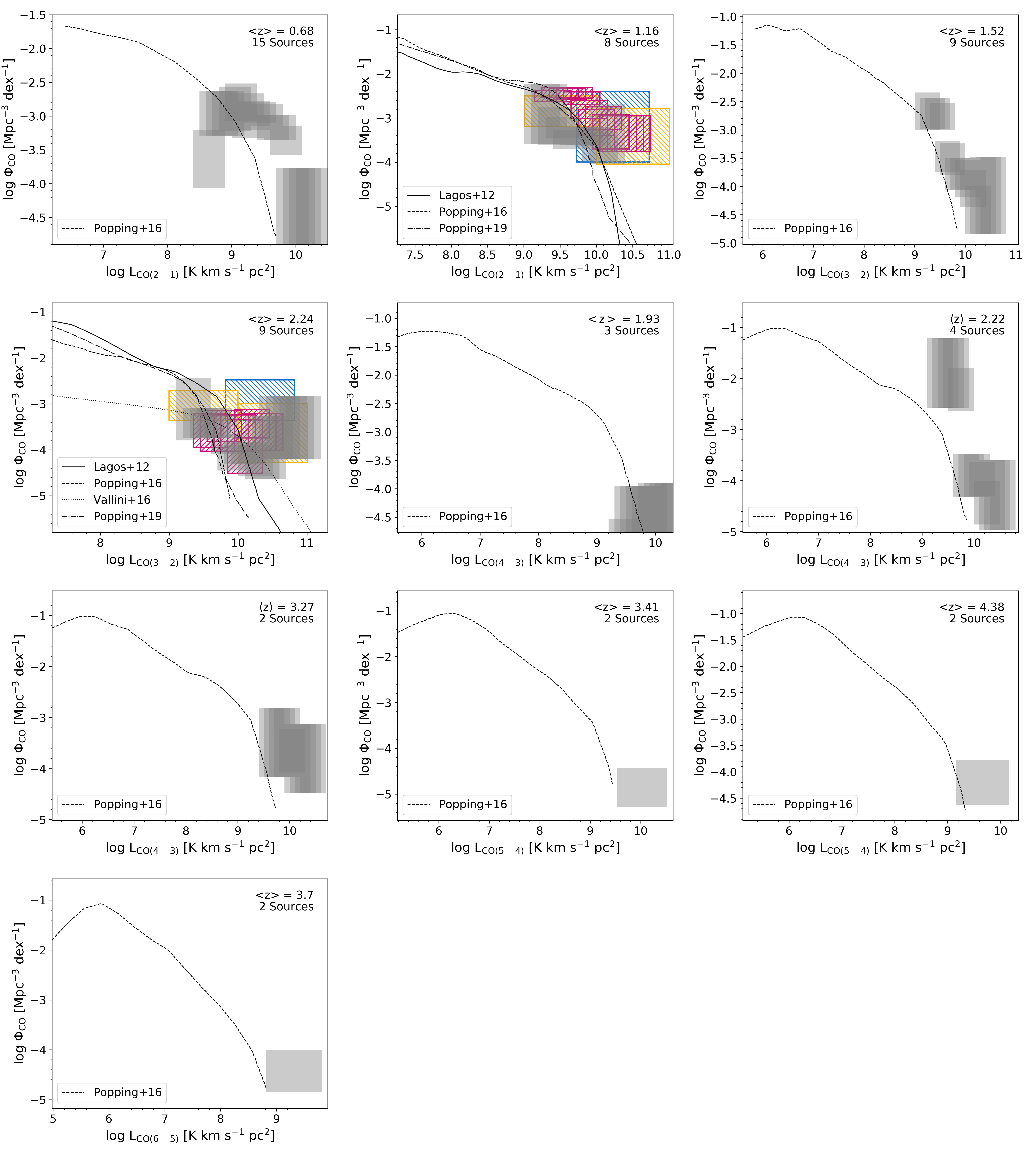}
    \caption{The PHIBSS2 CO luminosity functions observed here (shaded gray boxes, with sizes corresponding to 1$\sigma$ uncertainties), compared to the PdBI HDF-N work \citep[blue left-hatched boxes;][]{walter14}, the ASPECS pilot work  \citep[yellow left-hatched boxes;][]{decarli16}, the ASPECS LP work \citep[magenta right-hatched boxes;][]{decarli19}, the predicted CO luminosity function of \citet{vallini16} based on the \textit{Herschel} IR luminosity function, and the theoretical predictions of \citet{lagos12} and \citet{popping16}. Our derived CO luminostiy functions are consistent with constraints from previous work, but are in tension with the semi-analytic model predictions, particularly at the higher-J CO transitions where we observe larger number densities at higher CO luminosities than is predicted by these models.}
    \label{fig:lfs}
\end{figure*}

Figure \ref{fig:lfs} plots the PHIBSS2 CO luminosity function for a range of CO transitions and median redshifts in gray shaded boxes. Our results are plotted as a moving average, by displacing each luminosity bin by 0.1 dex and recalculating the CO luminosity function according to equation \ref{eq:co_lf}. In each panel, we give the number of candidate sources used to derive the given CO luminosity function and their median redshift. We are able to constrain CO\,($2-1$) at $\langle z \rangle \sim 0.7$ and $1.2$, CO\,($3-2$) at $\langle z \rangle \sim 1.5$ and $2.2$, CO\,($4-3$) at $\langle z \rangle \sim 1.9$, $2.2$ and $3.3$, CO\,($5-4$) at $\langle z \rangle \sim 3.4$ and $4.4$, and CO\,($6-5$) at $\langle z \rangle \sim 3.7$. We compare each of these to existing theoretical predictions from \citet{popping19}\footnote{We convert the molecular hydrogen mass functions to CO luminosity functions assuming an $\alpha_{\mathrm{CO}} = 3.6$ M$_{\odot}$ (K\,km\,s$^{-1}$\,pc$^{2}$)$^{-1}$ and temperature ratios of $r_{\mathrm{J1}} = 0.76 \pm 0.09$, $0.42 \pm 0.07$ for J $=$ 2, 3 respectively \citep{daddi15}.}, \citet{popping16}, \citet{vallini16}, and \citet{lagos12} and where possible, to existing observational constraints. To be able to consistently compare with the work of \citet{walter14} and \citet{decarli16,decarli19}, we calculate our uncertainties on the CO luminosity function in the same way. Thus the error bars along the y-axis correspond to Poissonian errors on $N_{i}$, the number of sources within a luminosity bin $i$, at the 1$\sigma$ level according to Tables 1 and 2 of \citet{gehrels86}, while the ``error bars'' along the x-axis simply reflect the width of the luminosity bin.

We fit our observed CO luminosity functions with a Schechter function \citep{schechter76} in the logarithmic form used by \citet{riechers19} and \citet{decarli19}:

\begin{equation}
    \mathrm{log\,\Phi(}L'\mathrm{)} = \mathrm{log\,\Phi^{*}}\, + \, \alpha\,\mathrm{log} \left( \frac{L'}{L'^{*}} \right)\, -\, \frac{1}{\mathrm{ln}\,10} \frac{L'}{L'^{*}}\, +\, \mathrm{log(ln(10))}
\end{equation}

\noindent where $\Phi^{*}$ is the scale number of galaxies per unit volume, $L'^{*}$ is the scale line luminosity and the parameter that sets the ``knee" of the luminosity function, and $\alpha$ is the slope of the faint end. 

To obtain estimates of the allowed range of Schechter parameters, we  fit the characteristic parameters described above to our CO\,($2-1$) at $<z> = 0.68$ luminosity function due to the small numbers of sources in all other cases. To account for the uncertainties of each luminosity bin, we draw points from normal distributions centered in each luminosity bin, with standard deviation corresponding to the size of the luminosity bin. We fit a Schechter function to that set of points while assuming unconstrained priors on the characteristic Schechter parameters. We then repeat the process with a new set of randomly drawn points from each luminosity bin and do this until enough points have been drawn to determine the posterior likelihood distributions of each Schechter parameter.

\begin{figure}
    \centering
    \includegraphics[width=\columnwidth]{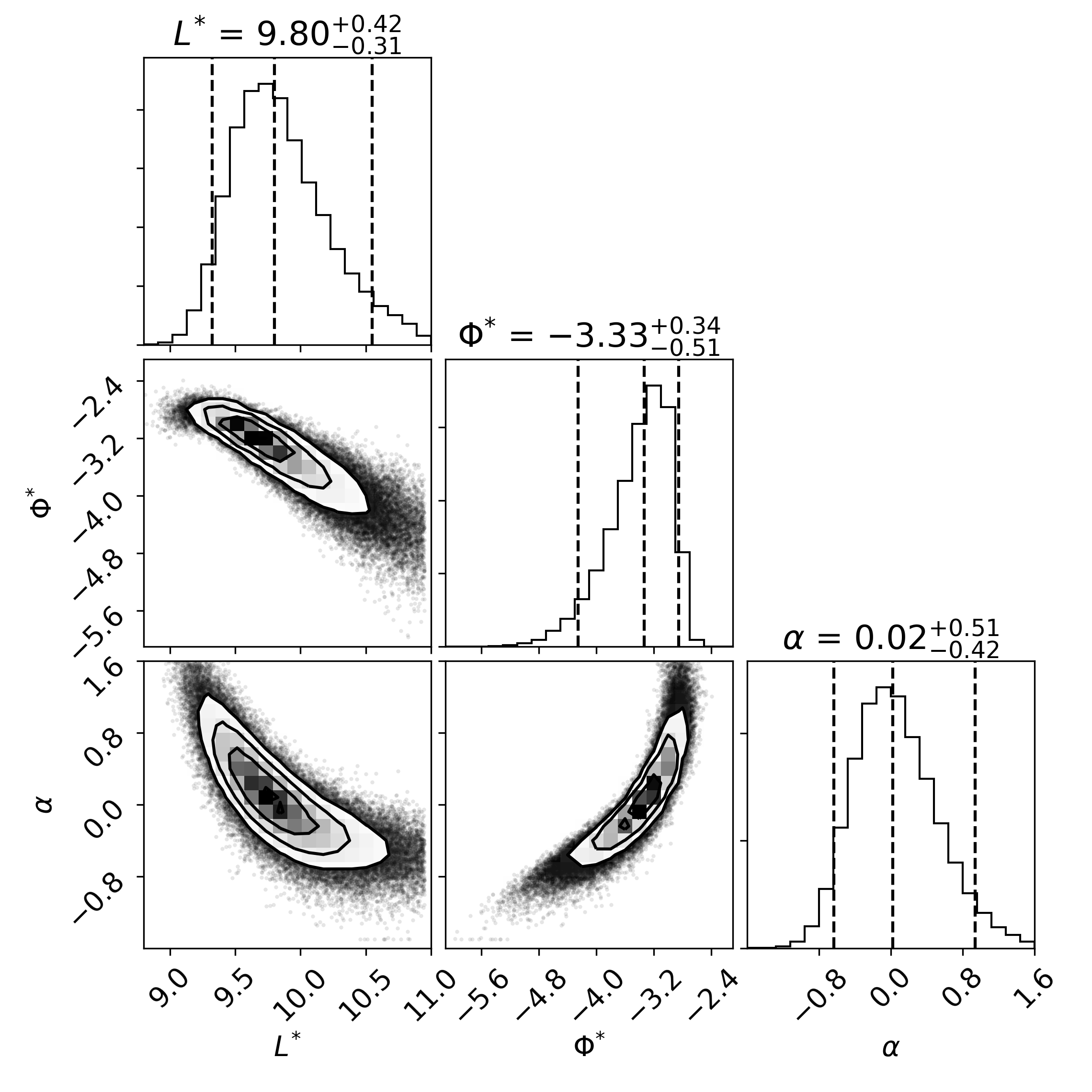}
    \caption{Corner plot of the Schechter model parameters posterior distribution from fitting the CO\,($2-1$) luminosity function. The parameters are reasonably well constrained by the data.}
    \label{fig:corner}
\end{figure}

\begin{figure}
    \centering
    \includegraphics[width=\columnwidth]{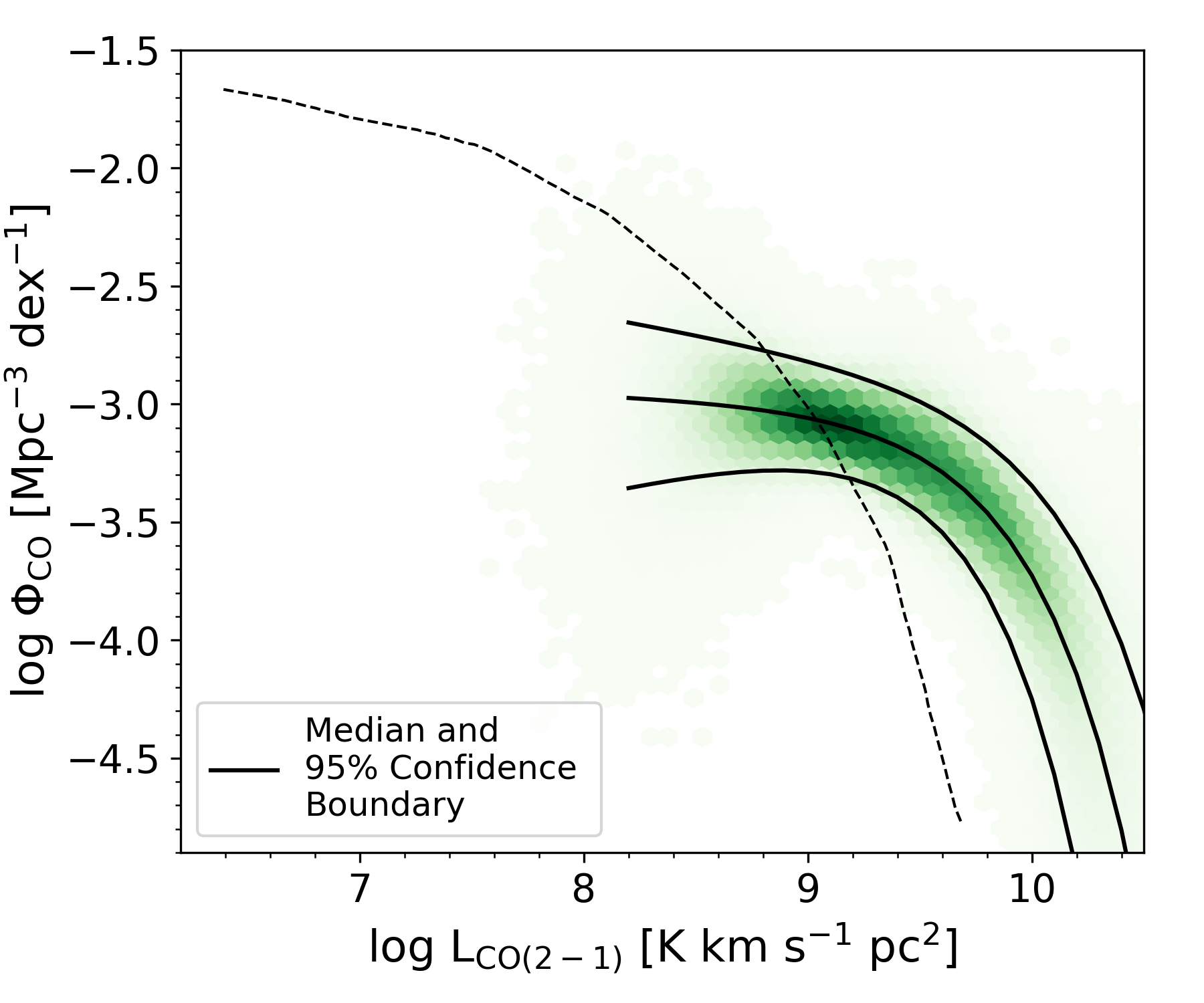}
    \caption{Density of Schechter fits for the CO\,($2-1$) $z \sim 0.7$ luminosity function. The black lines correspond to the median points and the boundary where 95\% of the fits lie. For reference, the \citet{popping16} prediction is plotted at the dashed black line. We see from this that the ``knee'' of the CO luminosity function is well constrained by the data, while there is more uncertainty in constraining the slope of the faint end.}
    \label{fig:schechter_hist}
\end{figure}

We show the results of this fitting in Figure \ref{fig:corner}, where we also include the posterior likelihood distribution of each parameter along with the 5th, 50th, and 95th percentiles. In Figure \ref{fig:schechter_hist}, we show the density of Schechter function fits to each sample of points drawn from the data. This Figure shows that the uncertainties are dominated by the faint end, below the ``knee" of the luminosity functions. However, the three parameters $L'^{*}_{\mathrm{CO}}$, $\Phi^{*}_{\mathrm{CO}}$, $\alpha$ are fairly reliably constrained by the data. We summarize the constraints on the Schechter model parameters for each fit in Table \ref{tab:schechter}, including the 5th and 95th percentiles.

\begin{deluxetable}{lcccc}
\tablecaption{Schechter Function Fit Parameter Constraints from PHIBSS2} \label{tab:schechter}
\tablewidth{700pt}
\tabletypesize{\footnotesize}
\tablehead{
	\colhead{Line} 			                &
	\colhead{Redshift}                      &
	\colhead{log $L'^{*}_{\mathrm{CO}}$}    &
	\colhead{log $\Phi^{*}_{\mathrm{CO}}$}  &
	\colhead{$\alpha$}                      
} 
\startdata
CO\,($2-1$) & 0.33 $-$ 0.99 & 9.76$^{+0.41}_{-0.31}$ & -3.31$^{+0.38}_{-0.58}$ & -0.07$^{+0.55}_{-0.45}$
\enddata
\end{deluxetable}

\subsection{Molecular Gas Mass Density Evolution}
To derive constraints on the evolution of co-moving molecular gas mass we need to convert our high-J CO luminosities to CO\,($1-0$) luminosities. We assume Rayleigh-Jeans brightness temperature ratios of $r_{\mathrm{J1}} = 0.76 \pm 0.09$, $0.42 \pm 0.07$, $0.31 \pm 0.06$,  and $0.23 \pm 0.04 $ for J = 2, 3, 4, and 5 respectively \citep{daddi15}. We then convert these CO\,($1-0$) luminosities to molecular gas masses, using an $\alpha_{\mathrm{CO}}$ value of 3.6 $\mathrm{M_{\odot}}$ ($\mathrm{K}$ $\mathrm{km}$ $\mathrm{s^{-1}}$ $\mathrm{pc^{2}}$)$^{-1}$ for the sake of consistency with previous work using

\begin{equation}
    M_{\mathrm{H_{2}}} = \alpha_{\mathrm{CO}}L'_{\mathrm{(CO\,(1-0))}}.
    \label{eq:gas_mass}
\end{equation}

The PHIBSS project has consistently used a 20\% larger value of $\alpha_{\rm CO}$ \citep{tacconi18}. \citet{carleton17} investigate the dependence of the conversion factor $\alpha_{\mathrm{CO}}$ on total mass surface density for $z > 1.7$ in the PHIBSS sample of galaxies and find that $92 - 100$\% of $\alpha_{\mathrm{CO}}$ measurements are similar to the Milky Way value of 4.36 M$_{\odot}$ (K\,km\,s$^{-1}$\,pc$^{2}$)$^{-1}$ adopted by PHIBSS. Here we use a value of 3.6 to compare consistently to other results in the literature who have adopted this value \citep{riechers19,decarli19}. Adopting a different constant value of $\alpha_{\mathrm{CO}}$ will result in a straightforward linear scaling of our M$_{\mathrm{H_{2}}}$ and $\rho(\mathrm{H_{2}})$ measurements.

\begin{figure*}[!ht]
\begin{center}
\includegraphics[width = \textwidth]{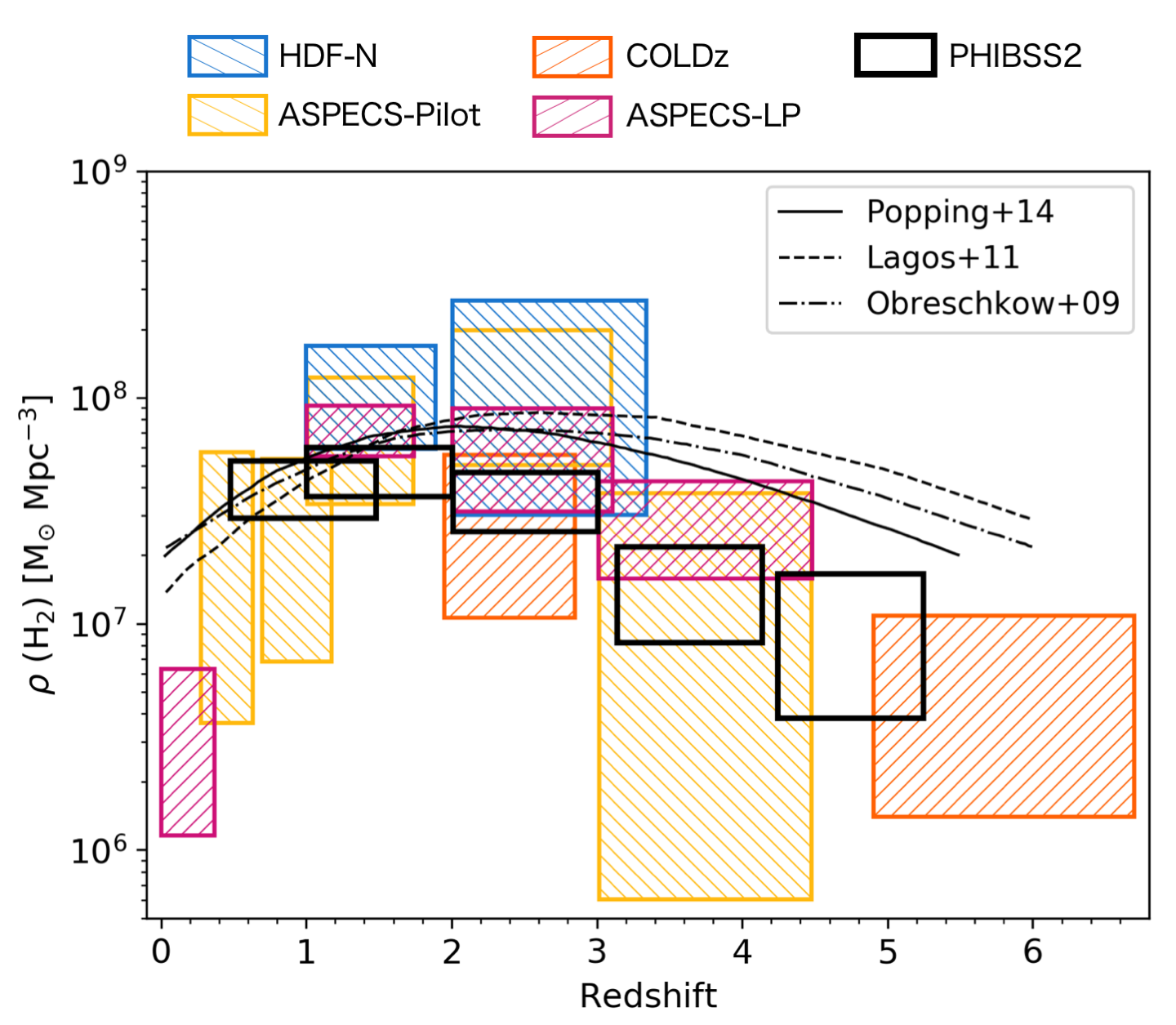}
\caption{The evolution of the molecular gas mass density with redshift, where the black boxes represent the constraints from the PHIBSS2 data. Orange right hatched boxes correspond to the constraints derived from the VLA COLDz measurements of \citet{riechers19}, purple right hatched boxes correspond to the constraints of ASPECS LP measurements of \citet{decarli19}, yellow left hatched boxes correspond to the work of \citet{decarli16}, and the blue left hatched boxes correspond to the constraints from the work of \citet{walter14}. The dashed lines correspond to model predictions for the evolution of the molecular gas mass density, as derived by \citet{obreschkow09}, \citet{lagos11}, and \citet{popping14a}, \citet{popping14b}. The constraints derived from serendipitous detections of CO in the PHIBSS2 fields are consistent with those of previous blind surveys.}
\label{fig:mgmd}
\end{center}
\end{figure*}

As in \citet{walter14}, \citet{decarli16}, \citet{riechers19}, and \citet{decarli19}, we do not extrapolate to the undetected faint end of the luminosity functions and only use actual candidate sources. We should note that this conversion from high-J CO transitions to molecular mass is increasingly uncertain as J increases: this is unavoidable as the excitation requirements become increasingly stringent, and so a diminishing fraction of the gas emits brightly in these transitions. The only way around this constraint is to directly observe the J=$1-0$ or $2-1$ transitions at high redshift, but that requires more powerful facilities than those in existence \citep[such as the ngVLA,][]{decarli18}. Our results are shown as black boxes in Figure \ref{fig:mgmd}, where each box corresponds to the combination of candidate sources observed at any transition in the given redshift range. 

\section{Discussion}\label{sec:discussion}
\subsection{Comparison to Previous Blind CO Surveys}
\subsubsection{Luminosity Functions}
The CO\,($2-1$) at $\langle z \rangle \sim 1.2$ and the CO\,($3-2$) at $\langle z \rangle \sim 2.3$ were previously constrained by \citet{walter14}, \citet{decarli16}, and \citet{decarli19}. The observational constraints from \citet{walter14} are the result of a blind CO survey in part of the \textit{Hubble} Deep Field North. \citet{decarli16} observed a $\sim 1$ arcmin region of the \textit{Hubble} Ultra Deep Field (UDF) with ALMA (the ASPECS pilot program), while \citet{decarli19} derive their constraints from the ASPECS Large Program. The redshift ranges for which we derive constraints from CO\,($2-1$) and CO\,($3-2$) are very similar to these previous works, so we directly compare our measurements to them. We see from Figure \ref{fig:lfs} that our results correspond to approximately the same luminosity bins as \citet{decarli16}, and are in agreement with their results. \citet{decarli16} report an excess of CO-bright galaxies in the UDF with respect to theoretical predictions, and our results confirm this for the galaxies we observe in the 3D-HST/CANDELS fields sampled by our PHIBSS2 data. This implies that galaxies in this redshift bin are more gas-rich than is currently predicted by theoretical models. 

\citet{riechers19} derive the CO\,($1-0$) luminosity function for a median redshift of $z = 2.4$ in the COLDz program. Our CO\,($3-2$) luminosity function is derived for a median redshift of $z \sim 2.3$. We consider this difference in redshift to be negligible and therefore compare to the COLDz measurements without any modifications. To compare these results then, it is only necessary to assume a line ratio between these two transitions. To convert our CO\,($3-2$) luminosities to CO\,($1-0$) we use r$_{\mathrm{31}} =  0.42 \pm 0.07$ from \citet{daddi15}. 

\begin{figure}
    \centering
    \includegraphics[width=\columnwidth]{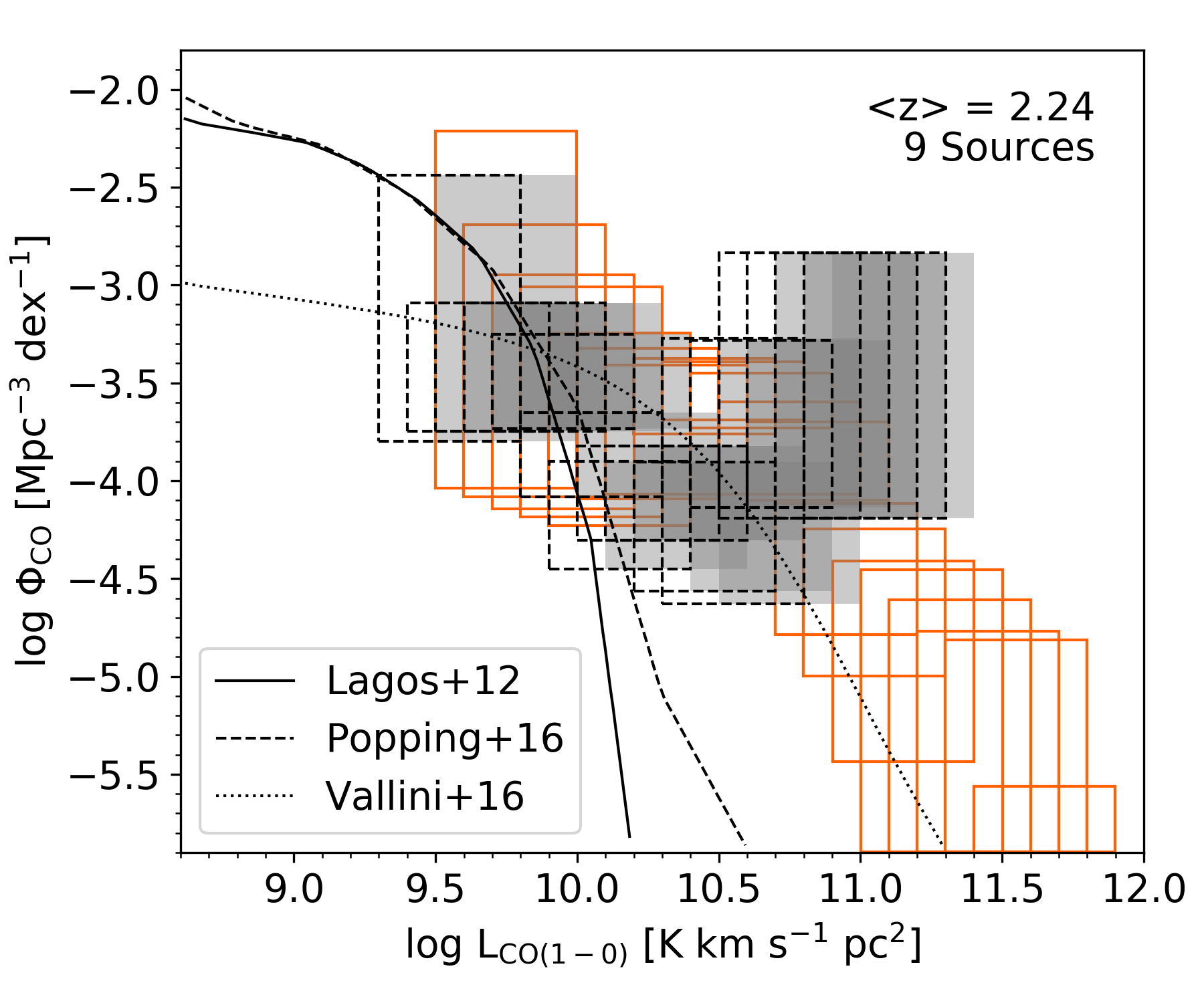}
    \caption{The comparison of our CO\,($3-2$) luminosity function, converted to CO\,($1-0$) assuming a brightness temperature ratio of $r_{31} = 0.42$ (gray boxes) to the results of \citet{riechers19} (orange boxes). Within the uncertainties, our measurements are consistent with those of \citet{riechers19}. As an additional comparison, we convert out CO\,($3-2$) luminosities with a $r_{31} = 0.66$ \citep[][black dashed boxes]{bolatto15}. This shifts our measurements to lower CO luminosities but our results stay in agreement with \citet{riechers19}. We plot the predictions of \citet{lagos12,vallini16,popping16} as a reference.}
    \label{fig:pavesi}
\end{figure}

We show the comparison between our derivation and that of the COLDz results of \citet{riechers19} in Figure \ref{fig:pavesi}. Overall we find that our measurements are consistent with those of \citet{riechers19} within the uncertainties, although there is a hint that our results may point to higher number densities than those measured in COLDz. This could be due to cosmic variance (CV), or it could also be evidence that higher-J observations or surveys tend to preferentially select higher gas excitation galaxies. This would then mean that our temperature ratio is too low. \citet{bolatto15} find for two CO\,($3-2$) bright $z \sim 2.2 -2.3$ galaxies $r_{31}$ ratios of order unity, while samples of nearby galaxies, luminous infrared galaxies, and ultra luminous infrared galaxies show mean values of $r_{31} \sim 0.66$. We convert our CO\,($3-2$) luminosities to CO\,($1-0$) using this higher temperature ratio and show the result in Figure \ref{fig:pavesi} (dashed black boxes). The change in assumed excitation produces a moderate shift toward lower luminosities, which brings the data into somewhat better agreement but does not completely eliminate the tension between both sets of measurement.

\subsubsection{Molecular Gas Mass Density Evolution}
In Figure \ref{fig:mgmd}, we compare our results to all previous observational constraints: those from \citet{walter14}, the ASPECS pilot work of \citet{decarli16}, the COLDz measurements of \citet{riechers19}, and the ASPECS LP measurements of \citet{decarli19}. Within the uncertainties, our results are consistent with all previous observational constraints. Between redshifts of $z \sim 2-3$, our result is most consistent with the measurement of \citet{walter14} and the ASPECS pilot, and hints at maybe a higher molecular gas mass density than that obtained by COLDz. From redshifts of $z \sim 3-5$, our measurements are consistent with the ASPECS pilot measurements, and hints at a lower molecular gas mass density than derived in the ASPECS-LP. Given the present state of the art in the uncertainties it is unclear if these discrepancies are real, but their magnitude is easily explained by cosmic variance.

\subsection{Cosmic Variance}
To address the question of cosmic variance, we use the results of \citet{driver10} to quantify the cosmic variance of the PHIBSS2 data. The authors repeatedly extract galaxy counts in cells of fixed size at random locations in the Sloan Digital Sky Survey (SDSS) Data Release 7. They explore the variance of the SDSS data in square cells from 1 to 2048 square degrees and in rectangular cells with aspect ratios ranging from 1:1 to 1:128. They find that cosmic variance depends on total survey volume, the survey aspect ratio, and whether the survey area is contiguous or composed on independent lines of sight, with cosmic variance decreasing for higher aspect ratios and non-contiguous survey areas (which essentially help sample a larger range of environments). 
\citet{driver10} provide a general equation that can be used at any redshift to estimate the cosmic variance of a survey \citep[eq. 4 in][]{driver10}:
\begin{multline} \label{eq:cos_var}
    \zeta_{\mathrm{Cos.Var.}} (\%) = (1.00 - 0.03\sqrt{(A/B)-)} \\ \times [219.7-52.4 \:\mathrm{log_{10}}(AB\times291.0) \\ 
    +  3.21[\:\mathrm{log_{10}}(AB\times 291.0)]^{2}]/\sqrt{NC/291.0}
\end{multline}

\noindent where $A$ and $B$ are the transverse lengths at the median redshift, $C$ is the radial depth all expressed in units of $h_{0.7}^{-1}$ Mpc, and $N$ is the number of independent sightlines. This empirical expression for estimating the cosmic variance is implemented as a function in the {\tt R} library {\tt celestial}, as \texttt{cosvarsph}. We input into \texttt{cosvarsph} the RA, Dec, and redshift limits that correspond to those values of the PHIBSS2 data cube with the median volume (since not all data cubes have the same size), and finally take $N$ to be the total number of data cubes. With this estimate, we derive cosmic variances in the range $\sim 13\% - 18\%$; these values are summarized in the last column of Table \ref{tab:volumes}. 

For comparison, we perform a crude estimate of the cosmic variance in the COLDz survey. The COLDz survey covers an area of 8.9 arcmin$^{2}$ at 31 GHz and 7.0 arcmin$^{2}$ at 39 GHz for COSMOS and an area of 50.9 arcmin$^{2}$ at 30 GHz and 46.4 arcmin$^{2}$ at 38 GHz. Using the average area and the redshift limits reported in their Figure 1 for both the CO\,($1-0$) and CO\,($2-1$) transitions as inputs into \texttt{cosvararea}, we estimate a cosmic variance of $\sim 34$\% and $\sim 24$\% for COSMOS and GOODS-N respectively. Performing the same estimate for the ASPECS-LP using the redshift limits from Table 1 of \citet{decarli19}, we find that the cosmic variance in the range $\sim 59\%$ to $\sim35\%$ for CO\,($1-0$) to CO\,($4-3$). These estimates are for a square survey area with aspect ratio 1:1, which is only approximate for either the COLDz or the ASPECS-LP surveys, and are therefore likely upper limits on their cosmic variance. However, this still shows that cosmic variance may be less of an issue in surveys that are composed of multiple independent lines of sight rather than one contiguous area.

\section{Conclusions} \label{sec:conclusions}
We present a catalog of 67 candidate secondary sources observed in 110 observations of PHIBSS2, where the primary target is a known optical high-$z$ galaxy, which includes spectra, redshifts, line widths, integrated fluxes, CO luminosities, and molecular gas masses. We perform an analysis of the false positive probabilities for each candidate secondary source, characterizing them with a reliability parameter $R$,  and assess the completeness of the search algorithm. We perform a search for optical counterparts corresponding to each candidate source, taking into account the redshift uncertainty for the optical sources in the 3D-HST/CANDELS catalogs. We find that $\sim$ 64\% of these secondary detections have optical counterparts (in some cases more than one) and include these together with an estimate of the probability of association in our catalog. Finally, we use the catalog of candidate sources to build the CO\,($2-1$), CO\,($3-2$), CO\,($4-3$), CO\,($5-4$), and CO\,($6-5$) luminosity functions for a range of median redshifts, spanning $z \sim 0.6 - 3.6$ and a volume sampled of $\sim 13500 - 57000$ Mpc$^{3}$ depending on the CO transition. We find broad agreement between our results and those of \citet{walter14}, \citet{decarli16}, \citet{riechers19}, and \citet{decarli19}. We also demonstrate that a blind CO search across many independent fields in observations of targeted objects can be successfully combined to establish constraints on the luminosity functions of different CO transitions in different redshift bins. We show that, in the case of CO, there appears to be little or no bias towards physically associated neighbors of the primary target down to the luminosities probed. We use an estimate of the cosmic variance to show that an approach which combines multiple independent fields mitigates the impact of cosmic variance. This is because for a contiguous survey area, the volume sampled needs to be very large in order to cut across many different environments; on the order of 10$^{7}$ h$_{0.7}^{-3}$ Mpc$^{3}$ (for an aspect ratio of 1:1) to decrease cosmic variance to a 10\% level according to the formalism by \citet{driver10}. This approach also exploits existing data which can significantly expand blind survey samples. The caveat is that one must deal with non-uniform sensitivity, which can however be handled through a good SNR characterization of the data sets.

We have derived the molecular gas mass density evolution from converting our high-J CO luminosity functions to CO\,($1-0$), assuming a CO luminosity to molecular gas mass conversion factor of $\alpha_{\mathrm{CO}} = 3.6$ $\mathrm{M_{\odot}}$ ($\mathrm{K}$ $\mathrm{km}$ $\mathrm{s^{-1}}$ $\mathrm{pc^{2}}$)$^{-1}$ for consistency with previous studies, and find our results to be largely consistent with previous constraints on the evolution of the cosmic cold gas mass density.

\acknowledgments
This work made use of PHIBSS `The Plateau de Bure HIgh-z Blue Sequence Survey' (Tacconi et al. 2010). This research has made use of the VizieR catalog access tool, CDS, Strasbourg, France. The original description of the VizieR service was published in \citep{ochsenbein00}. This work is based on observations taken by the 3D-HST Treasury Program (GO 12177 and 12328) with the NASA/ESA HST, which is operated by the Association of Universities for Research in Astronomy, Inc., under NASA contract NAS5-26555. This research has made use of the NASA/IPAC Extragalactic Database (NED) which is operated by the Jet Propulsion Laboratory, California Institute of Technology, under contract with the National Aeronautics and Space Administration. This research made use of APLpy, an open-source plotting package for Python \citep{robitaille12}. We would like to thank the anonymous referee for helping to improve this work through thoughtful comments and feedback. L.L. would like to thank Petr Pokorny for helpful discussions throughout the entirety of this project. L.L. and A.D.B. acknowledge partial support from NSF-AST 1412419. A.D.B. also acknowledges visiting support from the Alexander von Humboldt Foundation. 


\software{APLpy \citep{robitaille12},
          Astropy \citep{astropy:2013,astropy:2018},
          EAZY \citep{brammer08},
          IPython \citep{PER-GRA:2007},
          matplotlib \citep{hunter07},
          \& NumPy \citep{oliphant2006guide}.
          }

\appendix
\section{Description of Individual Candidate Sources}\label{app:spec}
\begin{figure*}
    \centering
    \includegraphics[width=\textwidth]{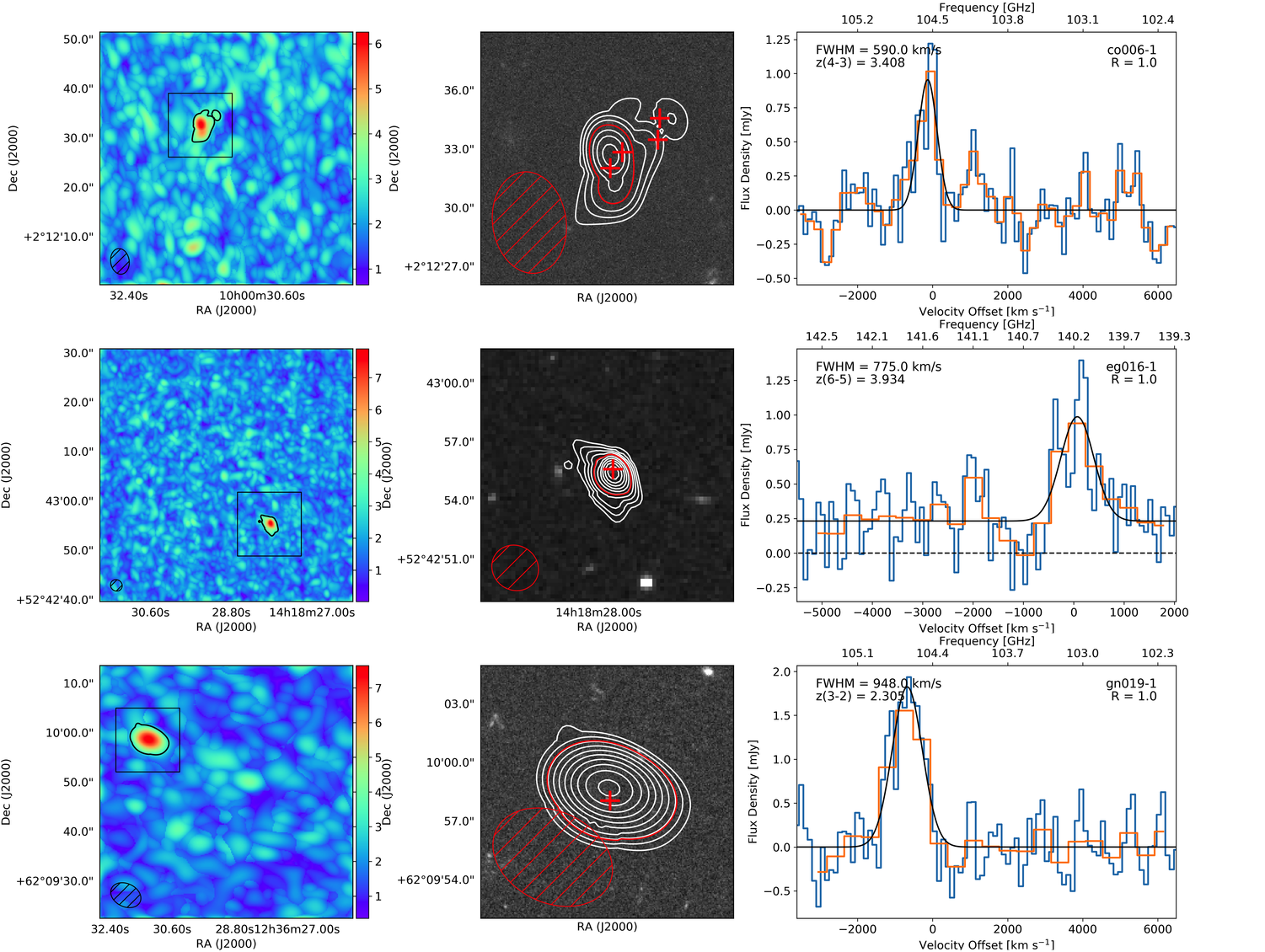}
    \caption{The brightest serendipitous CO sources in the COSMOS (top), EGS/AEGIS (middle), and GOODS-N (bottom) fields. The complete set of figures is available in the online journal.}
    \label{fig:sources}
\end{figure*}

Figure \ref{fig:sources} shows the brightest serendipitous CO sources in the COSMOS, EGS/AEGIS, and GOODS-N fields. The complete set of figures is available in the online journal. The left panels show the signal-to-noise maps at the velocity resolution where each source is detected with the highest SNR. The black contour corresponds to the detection threshold (largest negative SNR). The black box shows the zoom in region for the middle panel images, and the beam is shown in the bottom left corner. The middle panels are HST ACS F814W images where the white contours start at the 3$\sigma$ level and increase in steps of 0.5$\sigma$, the red contour corresponds to the detection threshold, red crosses mark the positions of tentative optical counterparts, and the beam is again shown in the bottom left corner. The size of the red contour appears small in some cases, this is however not a problem since these contours are really just the peak of the beam over our adopted ``threshold". The right panel shows the spectrum of each source extracted at the peak pixel, given unresolved sources. The blue spectrum corresponds to a velocity resolution of $\sim$ 100 km\,s$^{-1}$ while the orange spectrum corresponds to the velocity resolution matching that of the left panel. In cases where these two are the same, only the blue lines are shown. The FWHM and redshift of each candidate are added in the top left corners, and the source name (used in Tables \ref{tab:props} and \ref{tab:cntrprts}) and reliability measurement are included in the top right corner.

Table \ref{tab:props} lists each candidate source, divided according to the three fields (COSMOS, GOODS-N, EGS/AEGIS) with their RA and Dec, central frequency, flux, FWHM, SNR, and their completeness and reliability measures. Table \ref{tab:cntrprts} lists the potential optical counterparts with their IDs, RA and Dec, redshift, and angular separation. Redshifts extracted from the 3D-HST/CANDELS catalogs are photometric redshifts, except where otherwise noted.

\figsetstart
\figsetnum{1}
\figsettitle{Serendipitous CO Sources Spectra and Properties}
\figsetgrpstart
\figsetgrpnum{1.1}
\figsetgrptitle{}
\figsetplot{COS1.png}
\figsetgrpnote{}
\figsetgrpnum{1.2}
\figsetgrptitle{}
\figsetplot{COS2.png}
\figsetgrpnote{}
\figsetgrpnum{1.3}
\figsetgrptitle{}
\figsetplot{COS3.png}
\figsetgrpnote{}
\figsetgrpnum{1.4}
\figsetgrptitle{}
\figsetplot{COS4.png}
\figsetgrpnote{}
\figsetgrptitle{}
\figsetplot{COS5.png}
\figsetgrpnote{}
\figsetgrptitle{}
\figsetplot{COS6.png}
\figsetgrpnote{}
\figsetgrptitle{}
\figsetplot{GN1.png}
\figsetgrpnote{}
\figsetgrptitle{}
\figsetplot{GN2.png}
\figsetgrpnote{}
\figsetgrptitle{}
\figsetplot{GN3.png}
\figsetgrpnote{}
\figsetgrptitle{}
\figsetplot{GN4.png}
\figsetgrpnote{}
\figsetgrptitle{}
\figsetplot{GN5.png}
\figsetgrpnote{}
\figsetgrptitle{}
\figsetplot{GN6.png}
\figsetgrpnote{}
\figsetgrptitle{}
\figsetplot{EGS1.png}
\figsetgrpnote{}
\figsetgrptitle{}
\figsetplot{EGS2.png}
\figsetgrpnote{}
\figsetgrptitle{}
\figsetplot{EGS3.png}
\figsetgrpnote{}
\figsetgrpend
\figsetend

\startlongtable
\setcounter{table}{2}
\begin{deluxetable*}{lccccccccc}
\tablecaption{Properties of Potential detections \label{tab:props}}
\tablewidth{700pt}
\tabletypesize{\footnotesize}
\tablehead{
	\colhead{ID} 					& 
	\colhead{RA} 					& 
	\colhead{Dec} 					& 
	\colhead{Frequency}			 	&  
	\colhead{Flux} 					& 
	\colhead{FWHM} 					& 
	\colhead{SNR} 					& 
	\colhead{Completeness}			&
	\colhead{Reliability} 				\\ 
	\colhead{}				 		& 
	\colhead{(J2000.0)}			 	& 
	\colhead{(J2000.0)}	 			& 
	\colhead{[GHz]} 				& 
	\colhead{[Jy km s$^{-1}$]}	 	& 
	\colhead{[km s$^{-1}$]}	 		& 
	\colhead{}		 				& 
	\colhead{} 						& 
	\colhead{}						 
} 
\startdata
\cutinhead{COSMOS}
co006-1	& 10h0m31.44 & 2d12m32.51s & 104.582 & 0.599 $\pm$ 0.139 & 590 $\pm$ 104 & 6.26	& 0.8209 & 1.00	\\
co006-5 & 10h0m30.25 & 2d12m27.91s & 102.524 & 0.172 $\pm$ 0.066 & 141 $\pm$ 41 & 4.48 & 0.1973 & 0.35 \\
co012-1	& 10h0m44.78 & 2d33m26.80s & 135.900 & 0.613 $\pm$ 0.146 & 151 $\pm$ 27	& 6.25 & 0.7720	& 1.00	\\
co012-3	& 10h0m44.58 & 2d33m42.20s & 134.577 & 0.599 $\pm$ 0.199 & 277 $\pm$ 69	& 4.88 & 0.7510	& 0.83	\\
co018-1 & 10h0m58.71 & 1d45m53.60s & 142.039 & 0.341 $\pm$ 0.082 & 227 $\pm$ 41 & 6.68 & 0.9219 & 0.99 \\
co018-2 & 10h0m58.59 & 1d46m2.000s & 143.244 & 0.152 $\pm$ 0.047 & 119 $\pm$ 27 & 5.15 & 0.0968 & 0.76 \\
co027-2	& 10h0m44.16 & 2d07m00.93s & 107.631 & 0.323 $\pm$ 0.091 & 309 $\pm$ 66	& 5.50 & 0.4237	& 0.99	\\
xl53-1 & 10h0m28.36 & 2d15m49.28s & 132.169 & 0.315 $\pm$ 0.091 & 63 $\pm$ 14 & 5.88 & 0.0061 & 0.93 \\
xl53-2 & 10h0m28.99 & 2d16m9.28s & 133.094 & 0.304 $\pm$ 0.107 & 63 $\pm$ 17 & 5.38 & 0.0053 & 0.46 \\
co007-3	& 10h0m24.64 & 2d29m48.88s & 153.419 & 0.315 $\pm$ 0.100 & 167 $\pm$ 40	& 4.92 & 0.4791	& 0.91	\\
co007-2	& 10h0m24.52 & 2d29m43.48s & 153.643 & 0.216 $\pm$ 0.053 & 59  $\pm$ 11	& 5.23 & 0.1134	& 0.76	\\
xu53-5 & 10h0m40.93 & 2d23m27.44s & 153.676 & 0.392 $\pm$ 0.127 & 131 $\pm$ 32 & 4.54 & 0.2351 & 0.89 \\
xu53-6	& 10h0m40.11 & 2d23m39.04s & 151.627 & 0.637 $\pm$ 0.226 & 347 $\pm$ 93	& 4.39 & 0.2125	& 0.39	\\
xu53-7	& 10h0m39.35 & 2d23m03.24s & 151.257 & 0.782 $\pm$ 0.298 & 282 $\pm$ 81	& 4.37 & 0.9045	& 0.35	\\
xh53-2	& 10h1m10.03 & 2d30m04.90s & 136.249 & 0.696 $\pm$ 0.226 & 588 $\pm$ 144 & 4.58 & 0.3127 & 0.88	\\
co005-1 & 10h0m27.13 & 2d17m51.20s & 109.715 & 0.297 $\pm$ 0.110 & 155 $\pm$ 43 & 4.60 & 0.4130 & 0.86    \\
co005-2 & 10h0m28.76 & 2d17m32.80s & 111.352 & 0.260 $\pm$ 0.087 & 131 $\pm$ 33 & 4.34 & 0.3219 & 0.61    \\
co005-3 & 10h0m28.84 & 2d17m53.20s & 109.868 & 0.166 $\pm$ 0.068 & 91 $\pm$ 28  & 4.33 & 0.1388 & 0.59    \\
co004-1	& 10h0m39.88 & 2d20m45.64s & 136.209 & 0.432 $\pm$ 0.161 & 908 $\pm$ 255 & 4.37 & 0.9414 & 0.71	\\
xr53-2	& 10h1m43.05 & 2d06m55.24s & 154.097 & 0.903 $\pm$ 0.228 & 358 $\pm$ 68 & 4.72	& 0.0095 & 0.69	\\
xr53-4	& 10h1m40.33 & 2d07m07.24s & 152.109 & 1.395 $\pm$ 0.366 & 766 $\pm$ 152 & 4.33	& 1.0000 & 0.09	\\
co002-1	& 10h0m16.88 & 2d23m10.77s & 109.639 & 0.177 $\pm$ 0.053 & 93  $\pm$ 21	& 5.20 & 0.1609	& 0.63	\\
xg53-1  & 10h2m15.90 & 1d37m20.80s & 142.676 & 0.435 $\pm$ 0.130 & 119 $\pm$ 27 & 5.32 & 0.1321 & 0.62 \\
xw53-1 & 10h0m34.62 & 2d16m38.12s & 132.701 & 1.015 $\pm$ 0.339 & 382 $\pm$ 97 & 4.73 & 0.6225 & 0.60 \\
xn53-1 & 10h0m10.05 & 2d35m46.24s & 135.884 & 0.678 $\pm$ 0.221 & 130 $\pm$ 32 & 5.03 & 0.3668 & 0.31 \\
xj53-1 & 10h1m47.94 & 2d23m12.00s & 114.346 & 0.461 $\pm$ 0.160 & 69 $\pm$ 18 & 4.81 & 0.0150 & 0.08 \\
\cutinhead{GOODS-N}
gn019-1	&	12h36m31.31	&	62d09m58.47s	&	104.643	&	1.840 $\pm$ 0.247	&	948 $\pm$ 96	&	7.64	&	0.9921	&	1.00	\\
xc55-3  &   12h36m11.91 &   62d14m4.80s     &   130.324 &   1.289 $\pm$ 0.406   &   638 $\pm$ 152   &   4.56    &   0.6556    &   0.99    \\
xc55-2	&	12h36m12.19	&	62d14m15.40s	&	131.516	&	1.156 $\pm$ 0.530	&	595 $\pm$ 232	&	4.69	&	0.3219	&	0.85	\\
xc55-4  &   12h36m8.04  &   62d14m9.40s     &   129.997 &   0.660 $\pm$ 0.267   &   251 $\pm$ 77    &   3.85    &   0.4351    &   0.62    \\
gn010-6 &   12h36m42.68 &   62d12m2.73s     &   137.174 &   0.587 $\pm$ 0.255   &   642 $\pm$ 215   &   4.04    &   0.3992  &   0.96    \\
gn010-3 &   12h36m39.05 &   62d12m2.13s     &   136.516 &   0.444 $\pm$ 0.184   &   310 $\pm$ 107   &   4.22    &   0.7280  &   0.42    \\
gn030-2	&	12h36m22.27	&	62d10m19.82s	&	110.859	&	0.190 $\pm$ 0.056	&	117 $\pm$ 26	&	5.32	&	0.1141	&	0.92	\\
gn030-3	&	12h36m26.67	&	62d10m15.62s	&	110.809	&	0.148 $\pm$ 0.055	&	90 $\pm$ 26	&	5.16	&	0.0420	&	0.36	\\
xg55-2	&	12h37m1.87	&	62d14m06.60s	&	151.712	&	0.192 $\pm$ 0.105	&	76  $\pm$ 38	&	5.15	&	0.1196	&	0.91	\\
xg55-6  &   12h37m2.73  &   62d14m26.60s    &   151.712 &   0.277 $\pm$ 0.111   &   156 $\pm$ 47    &   4.13    &   0.1225    &   0.13    \\
xa55-2 &    12h36m59.75 &   62d15m9.80s     &   132.414 &   0.286 $\pm$ 0.098   &   28 $\pm$ 8  &   4.95    &   0.0102    &   0.62    \\
gn001-3	&	12h37m26.27	&	62d20m39.29s	&	103.518	&	0.468 $\pm$ 0.191	&	773 $\pm$ 238	&	4.17	&	0.3312	&	0.56	\\
xf55-1	&	12h35m56.17	&	62d10m46.20s	&	142.516	&	0.160 $\pm$ 0.173	&	90 $\pm$ 70	&	4.98	&	0.0907	&	0.52	\\
gn011-1 &   12h37m19.35 &   62d18m50.56s    &   137.386 &   0.543 $\pm$ 0.147   &   136 $\pm$ 28    &   4.85    &   0.3091  & 0.51  \\
gn032-1	&	12h37m16.23	&	62d15m31.12s	&	148.202	&	0.380 $\pm$ 0.099	&	206 $\pm$ 41	&	5.19	&	0.4593	&	0.49	\\
gn018-2	&	12h36m32.26	&	62d16m03.68s	&	129.890	&	0.532 $\pm$ 0.225	&	642 $\pm$ 205	&	4.44	&	0.2299	&	0.43	\\
gn018-3	&	12h36m31.74	&	62d16m00.48s	&	128.524	&	0.595 $\pm$ 0.185	&	637 $\pm$ 150	&	4.35	&	0.3020	&	0.27	\\
gn037-2	&	12h36m25.23	&	62d10m41.60s	&	226.683	&	0.178 $\pm$ 0.049	&	65  $\pm$ 14	&	5.55	&	0.0113	&	0.43	\\
gn006-1	&	12h36m32.72	&	62d17m48.71s	&	136.200	&	0.291 $\pm$ 0.091	&	63  $\pm$ 14	&	5.32	&	0.0130	&	0.41	\\
gn007-2	&	12h36m35.41	&	62d07m26.53s	&	145.737	&	0.431 $\pm$ 0.123	&	154 $\pm$ 34	&	5.29	&	0.3004	&	0.41	\\
gn034-2	&	12h36m20.86	&	62d19m07.70s	&	151.152	&	0.538 $\pm$ 0.188	&	383 $\pm$ 101	&	4.61	&	0.4066	&	0.39	\\
gn002-1	&	12h36m41.56	&	62d17m01.56s	&	114.944	&	0.550 $\pm$ 0.158	&	164 $\pm$ 35	&	4.92	&	0.4009	&	0.37	\\
gn021-3 &   12h36m0.69  &   62d11m26.18s    &   142.145 &   0.303 $\pm$ 0.095   &   58 $\pm$ 14     &   5.40    &   0.0337    &   0.37    \\
gn005-1	&	12h37m18.40	&	62d12m39.04s	&	101.340	&	0.235 $\pm$ 0.073	&	172 $\pm$ 42	&	5.29	&	0.3657	&	0.21	\\
gn017-6	&	12h36m53.43	&	62d17m04.86s	&	109.324	&	0.367 $\pm$ 0.132	&	237 $\pm$ 64	&	4.28	&	0.2002	&	0.20	\\
gn017-7	&	12h36m54.98	&	62d17m04.26s	&	106.419	&	0.320 $\pm$ 0.150	&	191 $\pm$ 70	&	4.27	&	0.4470	&	0.18	\\
gn017-1	&	12h36m53.78	&	62d17m15.26s	&	106.817	&	0.180 $\pm$ 0.056	&	83  $\pm$ 19	&	4.79	&	0.0950	&	0.09	\\
gn026-2	&	12h36m34.68	&	62d18m02.12s	&	106.883	&	0.430 $\pm$ 0.168	&	700 $\pm$ 206	&	4.27	&	0.3453	&	0.11	\\
\cutinhead{EGS}
eg016-1	&	14h18m27.95	&	52d42m55.28s	&	140.127	&	0.635 $\pm$ 0.158	&	791 $\pm$ 150	&	7.90	&	0.2339	&	1.00	\\
xb54-2	&	14h19m36.82	&	52d51m13.60s	&	137.437	&	0.669 $\pm$ 0.248	&	165 $\pm$ 47	&	5.27	&	0.3905	&	1.00	\\
xb54-3	&	14h19m37.57	&	52d50m53.40s	&	139.610	&	0.563 $\pm$ 0.207	&	171 $\pm$ 47	&	4.43	&	0.2380	&	0.96	\\
xd54-2  &   14h19m47.37 &   52d54m22.80s    &   132.142 &   0.365 $\pm$ 0.112   &   108 $\pm$ 25    &   5.19    &   0.1105    &   1.00    \\
eg014-1	&	14h20m31.63	&	52d59m20.86s	&	135.400	&	0.175 $\pm$ 0.049	&	69  $\pm$ 15	&	5.72	&	0.1315	&	0.82	\\
xi54-1	&	14h19m42.46	&	52d52m05.81s	&	172.316	&	0.744 $\pm$ 0.182	&	203 $\pm$ 37	&	5.35	&	0.2646	&	0.75	\\
eg012-1	&	14h19m51.28	&	52d51m10.86s	&	150.082	&	0.241 $\pm$ 0.080	&	48  $\pm$ 13	&	5.67	&	0.0066	&	0.69	\\
eg012-2	&	14h19m51.43	&	52d51m14.66s	&	148.704	&	0.622 $\pm$ 0.202	&	353 $\pm$ 87	&	4.69	&	0.4550	&	0.69	\\
eg006-3	&	14h18m44.84	&	52d43m30.66s	&	154.766	&	0.150 $\pm$ 0.055	&	49  $\pm$ 14	&	5.29	&	0.0419	&	0.32	\\
eg006-4	&	14h18m45.72	&	52d43m37.06s	&	155.775	&	0.184 $\pm$ 0.099	&	83  $\pm$ 36	&	5.00	&	0.0838	&	0.31	\\
eg006-6	&	14h18m48.01	&	52d43m03.46s	&	154.780	&	1.262 $\pm$ 0.343	&	829 $\pm$ 170	&	4.46	&	0.8294	&	0.25	\\
eg005-1	&	14h18m58.38	&	52d42m59.43s	&	108.805	&	0.141 $\pm$ 0.050	&	61  $\pm$ 18	&	5.12	&	0.0537	&	0.25	\\
eg018-3	&	14h19m40.06	&	52d51m38.39s	&	206.825	&	0.366 $\pm$ 0.122	&	216 $\pm$ 54	&	4.96	&	0.8813	&	0.15	\\
\enddata
\end{deluxetable*}

\startlongtable
\begin{deluxetable*}{lccccccc}
\tablecaption{Properties of Potential Optical Counterparts \label{tab:cntrprts}}
\tablewidth{700pt}
\tabletypesize{\footnotesize}
\tablehead{
	\colhead{ID} 				                &
	\colhead{Counterpart\tablenotemark{a}} 	    &
	\colhead{RA\tablenotemark{b}}		 	    &
	\colhead{DEC\tablenotemark{b}}			    &
	\colhead{Redshift\tablenotemark{b}}		    &
	\colhead{CO Redshift\tablenotemark{c}}      &
    \colhead{$\Delta r$\tablenotemark{d}}       &
    \colhead{Association\tablenotemark{e}}      \\
    \colhead{} 				                    &
	\colhead{} 	                                &
	\colhead{}		 	                        &
	\colhead{}			                        &
	\colhead{}		                            &
	\colhead{}                                  &
    \colhead{[$''$]}                            &
    \colhead{}                                  
}
\startdata
\cutinhead{COSMOS}
co006-1 & COS 2997 & 10h0m31.43 & 2d12m32.018s & 3.52 & 3.4084 & 0.5 & 0.98 \\
co012-1 & COS 33686 & 10h0m44.79 & 2d33m26.58s & 0.6961\tablenotemark{f} & 0.6964 & 0.2 & 0.99 \\
co012-3 & COS 33774 & 10h0m44.59 & 2d33m17.67s & 0.67 & 1.5695 & 1.6 & 0.70 \\
co018-1 & COS 485943 & 10h0m58.69 & 1d45m54.00s & 0.65 & 0.6231 & 0.5 & 0.98 \\
co027-2 & COS 1236431 & 10h0m44.27 & 2d7m2.28s & 0.95 & 1.1419 & 2.1 & 0.42 \\
co007-3 & COS 2345915 & 10h0m24.48 & 2d29m50.10s & 1.25 & 1.2539 & 2.7 & 0.53 \\
xu53-5 & COS 20811 & 10h0m40.88 & 2d23m26.84s & 1.27 & 1.2502 & 1.0 & 0.89 \\
xu53-6 & COS 21168 & 10h0m40.08 & 2d23m40.42s & 0.67 & 0.5204 & 1.5 & 0.74 \\
xu53-7 & COS 20188 & 10h0m39.28 & 2d23m2.68s & 1.84 & 2.0481 & 1.2 & 0.84 \\
xh53-2 & COS 2307212 & 10h1m9.89 & 2d30m4.82s & 0.67 & 0.6921 & 2.2 & 0.07 \\
co005-1 & COS 11546 & 10h0m26.94 & 2d17m48.65s & 1.15 & 1.1012 & 3.8 & 0.23 \\
co005-1 & COS 11595 & 10h0m26.88 & 2d17m49.77s & 1.34 & 1.1012 & 4.0 & 0.15 \\
co005-1 & COS 11549 & 10h0m26.96 & 2d17m48.09s & 0.94 & 1.1012 & 4.0 & 0.15 \\
co005-2 & [Capak2017] 1599172 & 10h0m28.70 & 2d17m30.89s & 3.11 & 3.1404 & 2.1 & 0.77\\
co005-3 & COS 111603 & 10h0m29.02 & 2d17m50.71s & 4.26 & 4.2451 & 3.7 & 0.28 \\
co002-1 & COS 20413 & 10h0m16.82 & 2d23m10.93s & 1.85 & 2.1539 & 1.0 & 0.89 \\
xw53-1 & COS 09632 & 10h0m34.57 & 2d16m38.06s & 1.69 & 1.6058 & 0.8 & 0.88 \\
xn53-1 & COS 2361883 & 10h0m9.83 & 2d35m46.86s & 0.73 & 0.6966 & 3.4 & 0.08 \\
\cutinhead{GOODS-N}
gn019-1 & GN 05359 & 12h36m31.29 & 62d9m58.04s & 2.3301\tablenotemark{g} & 2.3045 & 0.5 & 0.99 \\
xc55-3 & GN 18914 & 12h36m11.79 & 62d14m7.69s & 1.99 & 1.6534, 2.5377 & 3.0 & 0.52 \\
xc55-3 & GN 18951 & 12h36m11.86 & 62d14m8.36s  & 2.07 & 0.7690, 1.6534, 2.5377 & 3.6 & 0.30 \\
xc55-2 & GN 19484 & 12h36m12.44 & 62d14m17.49s & 3.83 & 3.3817, 4.2577 & 2.7 & 0.61 \\
gn010-6 & GN 11844 & 12h36m42.78 & 62d12m3.97s & 1.70 & 1.5209 & 1.5 & 0.64 \\
gn010-6 & GN 11803 & 12h36m42.91 & 62d12m3.46s & 2.26 & 1.5209, 2.3610 & 1.8 & 0.48 \\
gn010-6 & GN 11666 & 12h36m42.49 & 62d12m1.21s & 1.65 & 0.6806, 1.5209, 2.3610 & 2.0 & 0.36 \\
xg55-2 & GN 18857 & 12h37m1.50 & 62d14m7.10s & 3.01 & 2.7984 & 2.7 & 0.20 \\
xg55-2 & GN 19030 & 12h37m1.85 & 62d14m9.47s & 0.46 & 0.5196 & 2.9 & 0.08 \\
xa55-2 & GN 22587 & 12h36m59.50 & 62d15m12.37s & 0.67 & 0.7595 & 3.1 & 0.32 \\
gn001-3 & GN 36945 & 12h37m26.17 & 62d20m39.54s & 1.00 & 1.2270 & 0.8 & 0.96 \\
gn018-2 & GN 25624 & 12h36m32.24 & 62d16m3.10s & 3.20 & 3.4366 & 0.6 & 0.91 \\
gn018-2 & GN 25623 & 12h36m32.07 & 62d16m2.86s & 0.42 & 0.7749 & 1.5 & 0.42 \\
gn018-3 & GN 25587 & 12h36m31.77 & 62d16m1.81s & 1.46 & 0.7937 & 1.3 & 0.57 \\
gn006-1 & [BIO2015] GNDV 6325117508 & 12h36m32.51 & 62d17m50.80 & 4.65 & 4.9225 & 1.7 & 0.59 \\
gn002-1 & GN 28898 & 12h36m41.28 & 62d17m3.17s & 2.31 & 2.0084 & 2.6 & 0.62 \\
gn002-1 & GN 28704 & 12h36m41.89 & 62d16m59.95s & 1.28 & 1.0056, 2.0084 & 2.8 & 0.56 \\
gn005-1 & GN 13860 & 12h37m18.08 & 62d12m39.43s	 & 1.27 & 1.2749 & 2.3 & 0.69 \\
gn017-6 & GN 28920 & 12h36m53.66 & 62d17m4.33s & 2.28 & 2.1630 & 1.7 & 0.87 \\
gn017-7 & GN 29041 & 12h36m55.07 & 62d17m5.93s	& 2.11 & 2.2494 & 1.8 & 0.85 \\
gn017-7 & GN 28807 & 12h36m55.02 & 62d17m1.54s & 0.70 & 1.1663, 2.2494 & 2.7 & 0.67 \\
gn017-1 & GN 1308 & 12h36m54.11 & 62d17m12.12s & 2.52 & 2.2373 & 3.9 & 0.31 \\
gn026-2 & GN 31722 & 12h36m34.99 & 62d18m1.07s & 1.67 & 1.1569, 2.2353 & 2.4 & 0.31 \\
\cutinhead{EGS}
eg016-1 & EGS 33606 & 14h18m27.92 & 52d42m55.59s & 4.16 & 3.9337 & 0.4 & 0.95 \\
xb54-2 & EGS 16871 & 14h19m36.89 & 52d51m14.44s & 0.86 & 0.6774 & 1.0 & 0.93 \\
xb54-3 & [Barro2011] 124112 & 14h19m37.37 & 52d50m50.32s & 0.68 & 0.6513 & 3.6 & 0.15 \\
xd54-2 & EGS 21954 & 14h19m47.52 & 52d54m21.54s & 1.66 & 1.6168 & 1.9 & 0.69 \\
eg014-1 & EGS 10403 & 14h20m31.70 & 52d59m22.04s & 1.74 & 1.5539 & 1.4 & 0.11 \\
eg012-1 & EGS 06254 & 14h19m51.40 & 2d51m9.70s & 2.04 & 2.0719 & 1.6 & 0.55 \\
eg012-1 & EGS 06217 & 14h19m51.36 & 52d51m8.85s & 1.62 & 1.3040, 2.0719 & 2.1 & 0.22 \\
eg012-1 & EGS 06392 & 14h19m51.46 & 52d51m12.31s & 0.94 & 0.5361, 1.3040 & 2.2 & 0.15 \\
eg006-3 & EGS 23517 & 14h18m44.64 & 52d43m32.71s & 3.39 & 3.4678 & 2.7 & 0.16 \\
eg006-4 & EGS 23149 & 14h18m45.56 & 52d43m37.29s & 1.18 & 0.4799 & 1.4 & 0.78 \\
eg006-6 & EGS 18930 & 14h18m48.11 & 52d43m3.12s & 1.4157\tablenotemark{g} & 1.2341, 1.9787 & 1.8 & 0.63 \\
eg005-1 & EGS 11013 & 14h18m58.51 & 52d43m0.15s & 1.80 & 1.1188, 2.1781 & 1.4 & 0.92 \\
eg018-3 & EGS 16186 & 14h19m40.07 & 52d51m39.71s & 1.48 & 0.6719, 1.2291, 1.7863 & 1.3 & 0.02 \\
\enddata
\tablenotetext{a}{Designation of the optical counterpart in the COSMOS, GOODS-N, and EGS/AEGIS catalogs.}
\tablenotetext{b}{RA, DEC, and redshift (``best") of the optical counterpart taken from the COSMOS, GOODS-N, and EGS/AEGIS catalogs.}
\tablenotetext{c}{List of redshifts corresponding to the possible CO transitions given the posterior likelihood distributions of the EAZY SED fitting.}
\tablenotetext{d}{Projected angular separation between candidate source and the potential optical counterpart.}
\tablenotetext{e}{For cases where multiple potential optical counterparts exist, we assign a probability of association that is proportional to the inverse square of the projected angular separation.}
\tablenotetext{f}{Spectroscopic redshift.}
\tablenotetext{g}{Grism redshift.}
\end{deluxetable*}

\section{Luminosity Function and Molecular Gas Mass Density} Constraints: Tabulated Results
In this appendix, we include the 1$\sigma$ ranges for each luminosity bin, for every CO luminosity function we measure, as shown in Figure \ref{fig:lfs}. Bins are 0.5 dex wide and given in steps of 0.1 dex, therefore every 5th bin is statistically independent. We also include the molecular gas mass density constraint for each redshift bin in tabulated form.

\startlongtable
\begin{deluxetable*}{cc|cc|cc}
\tablecaption{Luminosity functions of the observed CO transitions \label{tab:lfs}}
\tablewidth{700pt}
\tabletypesize{\small}
\tablehead{
	\colhead{log(L$'_{\mathrm{CO}}$) bin} 		& 
	\colhead{log $\Phi$, 1$\sigma$} 			&
	\colhead{log(L$'_{\mathrm{CO}}$) bin} 		& 
	\colhead{log $\Phi$, 1$\sigma$}             &
	\colhead{log(L$'_{\mathrm{CO}}$) bin} 		& 
	\colhead{log $\Phi$, 1$\sigma$}             \\ 
	\colhead{(K\,km\,s$^{-1}$\,pc$^{-2}$)}		& 
	\colhead{(Mpc$^{-3}$\,dex$^{-1}$)}			&
	\colhead{(K\,km\,s$^{-1}$\,pc$^{-2}$)}		& 
	\colhead{(Mpc$^{-3}$\,dex$^{-1}$)}			&
	\colhead{(K\,km\,s$^{-1}$\,pc$^{-2}$)}		& 
	\colhead{(Mpc$^{-3}$\,dex$^{-1}$)}			\\ 
} 
\startdata
\multicolumn{2}{c}{CO(2-1), z $\sim 0.34 - 0.79$} &
\multicolumn{2}{c}{CO(2-1), z $\sim 1.01 - 1.27$} &
\multicolumn{2}{c}{CO(3-2), z $\sim 1.22 - 1.66$} \\
8.4$-$8.9 & -4.06 , -3.21 & 9.0$-$9.5 & -3.59 , -2.23 & 9.0$-$9.5 & -3.00 , -2.34 \\
8.5$-$9.0 & -3.29 , -2.63 & 9.1$-$9.6 & -3.59 , -2.23 & 9.1$-$9.6 & -2.99 , -2.44 \\
8.6$-$9.1 & -3.29 , -2.63 & 9.2$-$9.7 & -3.60 , -2.75 & 9.2$-$9.7 & -2.99 , -2.44 \\
8.7$-$9.2 & -3.29 , -2.63 & 9.3$-$9.8 & -3.46 , -2.98 & 9.3$-$9.8 & -3.00 , -2.52 \\
8.8$-$9.3 & -3.22 , -2.56 & 9.4$-$9.9 & -3.71 , -3.27 &	9.4$-$9.9 & -3.74 , -3.19 \\
8.9$-$9.4 & -3.07 , -2.52 & 9.5$-$10.0 & -3.61 , -3.21 & 9.5$-$10.0 & -3.79 , -3.24 \\
9.0$-$9.5 & -3.32 , -2.66 & 9.6$-$10.1 & -3.61 , -3.21 & 9.6$-$10.1 & -4.05 , -3.50 \\
9.1$-$9.6 & -3.21 , -2.77 & 9.7$-$10.2 & -3.64 , -3.21 & 9.7$-$10.2 & -4.05 , -3.50 \\
9.2$-$9.7 & -3.10 , -2.79 & 9.8$-$10.3 & -3.98 , -3.32 & 9.8$-$10.3 & -4.19 , -3.54 \\
9.3$-$9.8 & -3.14 , -2.82 & 9.9$-$10.4 & -4.05 , -2.69 & 9.9$-$10.4 & -4.37 , -3.71 \\
9.4$-$9.9 & -3.35 , -2.98 & -- & -- & 10.0$-$10.5 & -4.84 , -3.98 \\
9.5$-$10.0 & -3.35 , -2.98 & -- & -- & 10.1$-$10.6 & -4.84 , -3.48 \\
9.6$-$10.1 & -3.57 , -3.14 & -- & -- & 10.2$-$10.7 & -4.84 , -3.48 \\
9.7$-$10.2 & -5.12 , -3.76 & -- & -- & 10.3$-$10.8 & -4.84 , -3.48 \\
9.8$-$10.3 & -5.12 , -3.76 & -- & -- & -- & -- \\
9.9$-$10.4 & -5.12 , -3.76 & -- & -- & -- & -- \\
10.0$-$10.5 & -5.12 , -3.76 & -- & -- & -- & -- \\
\hline
\multicolumn{2}{c}{CO(3-2), z $\sim 2.01 - 2.31$} &
\multicolumn{2}{c}{CO(4-3), z $\sim 1.03 - 1.98$} &
\multicolumn{2}{c}{CO(4-3), z $\sim 2.05 - 2.59$} \\
9.1$-$9.6 & -3.80,-2.44 &	8.8$-$9.3 & -8.32,-6.96 &	9.1$-$9.6 & -2.57,-1.21 \\
9.2$-$9.7 & -3.75,-3.09 &	8.9$-$9.4 & -8.32,-6.96 &	9.2$-$9.7 & -2.57,-1.21 \\
9.3$-$9.8 & -3.75,-3.09 &	9.0$-$9.5 & -8.32,-6.96 &	9.3$-$9.8 & -2.57,-1.21 \\
9.4$-$9.9 & -3.75,-3.09 &	9.1$-$9.6 & -8.32,-6.96 &	9.4$-$9.9 & -2.57,-1.21 \\
9.5$-$10.0 & -3.73,-3.25 &	9.2$-$9.7 & -5.38,-4.53 &	9.5$-$10.0 & -2.64,-1.79 \\
9.6$-$10.1 & -4.08,-3.65 &	9.3$-$9.8 & -5.31,-3.95 &	9.6$-$10.1 & -4.33,-3.47 \\
9.7$-$10.2 & -4.45,-3.90 &	9.4$-$9.9 & -5.31,-3.95 &	9.7$-$10.2 & -4.33,-3.47 \\
9.8$-$10.3 & -4.31,-3.82 &	9.5$-$10.0 & -5.31,-3.95 &	9.8$-$10.3 & -4.33,-3.47 \\
9.9$-$10.4 & -4.31,-3.82 &	9.6$-$10.1 & -5.31,-3.95 &	9.9$-$10.4 & -4.29,-3.64 \\
10.0$-$10.5 & -4.56,-3.91 &	9.7$-$10.2 & -5.25,-3.90 &	10.0$-$10.5 & -4.86,-4.01 \\
10.1$-$10.6 & -4.63,-3.27 &	9.8$-$10.3 & -5.25,-3.90 &	10.1$-$10.6 & -4.96,-3.60 \\
10.2$-$10.7 & -4.63,-3.27 &	9.9$-$10.4 & -5.25,-3.90 &	10.2$-$10.7 & -4.96,-3.60 \\
10.3$-$10.8 & -4.19,-2.83 &	10.0$-$10.5 & -5.25,-3.90 &	10.3$-$10.8 & -4.96,-3.60 \\
10.4$-$10.9 & -4.19,-2.83 &	10.1$-$10.6 & -5.25,-3.90 &	-- & -- \\
10.5$-$11.0 & -4.19,-2.83 &	-- & -- & -- & -- \\	
10.6$-$11.1 & -4.19,-2.83 &	-- & -- & -- & -- \\	
10.7$-$11.2 & -4.19,-2.83 &	-- & -- & -- & -- \\
\hline
\multicolumn{2}{c}{CO(4-3), z $\sim 3.14 - 3.40$} &
\multicolumn{2}{c}{CO(5-4), z $\sim 3.38 - 3.44$} &
\multicolumn{2}{c}{CO(5-4), z $\sim 4.25 - 4.51$} \\
9.4$-$9.9 & -4.17,-2.81 &	9.5$-$10.5 & -5.28,-4.43 &	9.2$-$10.2 & -4.62,-3.77 \\
9.5$-$10.0 & -4.17,-2.81 & -- & -- & -- & -- \\		
9.6$-$10.1 & -4.17,-2.81 & -- & -- & -- & -- \\		
9.7$-$10.2 & -4.17,-2.81 & -- & -- & -- & -- \\		
9.8$-$10.3 & -4.08,-3.22 & -- & -- & -- & -- \\		
9.9$-$10.4 & -4.48,-3.12 & -- & -- & -- & -- \\		
10.0$-$10.5 & -4.48,-3.12 & -- & -- & -- & -- \\		
10.1$-$10.6 & -4.48,-3.12 & -- & -- & -- & -- \\		
10.2$-$10.7 & -4.48,-3.12 & -- & -- & -- & -- \\		
\multicolumn{2}{c}{CO(6-5), z $\sim 3.47 - 3.93$} &
\multicolumn{2}{c}{} &
\multicolumn{2}{c}{} \\
8.8$-$9.8 & -4.85,-4.0 & & & & \\
\enddata
\end{deluxetable*}

\begin{deluxetable}{lc}
\tablecaption{Derived molecular gas mass density evolution constraints} \label{tab:h2_evolution}
\tablewidth{700pt}
\tabletypesize{\footnotesize}
\tablehead{
	\colhead{Redshift}                                     &
	\colhead{$\rho$(H2)}                                   \\
	\colhead{}                                             &
	\colhead{(1$\times$10$^{6}$ M$_{\odot}$ Mpc$^{-3}$)}
} 
\startdata
0.4799$-$1.4799  & 29.31$-$52.38 \\
1.0056$-$2.0056  & 36.55$-$59.98 \\
2.0084$-$3.0084  & 26.27$-$46.95 \\
3.1404$-$4.1404  & 8.22$-$21.76 \\
4.2451$-$5.2451  & 3.83$-$16.58 \\
\enddata
\end{deluxetable}

\bibliographystyle{aasjournal}
\bibliography{references}

\end{document}